%% file: main.tex
\documentclass[amsmath,trackchanges,twocolumn]{aastex702}
\usepackage{graphicx} 
\usepackage{booktabs}
\usepackage{amsmath,amssymb,tikz}
\usetikzlibrary{arrows.meta, fit, backgrounds}

\begin{document}
\title{The Maunder Model and Catalog: Stellar Rotation, Bimodal Activity, and Magnetic Braking in Kepler Main-Sequence Stars}


\author[orcid=0009-0008-5080-496X]{Ilay Kamai}
\affiliation{Physics Department, Technion: Israel Institute of Technology,
Haifa 32000,
Israel}
\email[show]{ilay.kamai@campus.technion.ac.il}  

\author[orcid=0000-0003-4897-2215]{Lavi Somers}
\affiliation{Physics Department, Technion: Israel Institute of Technology,
Haifa 32000,
Israel}
\email[show]{lavisomers@gmail.com}  

\author{Hagai B. Perets}
\email{hperets@physics.technion.ac.il}
\affiliation{Physics Department, Technion: Israel Institute of Technology,
Haifa 32000,
Israel}

\input{sections/abstract}
\input{sections/intro}
\input{sections/data}

\input{sections/model}
\input{sections/results}
\input{sections/catalog}
\input{sections/conclusions}

\bibliography{main}{}
\bibliographystyle{aasjournalv7.1}
\clearpage
\input{sections/appendix}

\end{document}

%% file: sections/abstract.tex
\begin{abstract}
We present The Maunder, a machine learning pipeline and resulting catalog of rotation periods for 148,746 main-sequence stars in the Kepler field. To overcome single-catalog systematics and the simulation-to-reality gap, our architecture employs a hybrid training objective: a joint-embedding self-supervised loss applied to all light curves, combined with a supervised loss trained strictly on cross-catalog consensus labels. By processing multi-scale time- and frequency-domain inputs over rolling windows, the model leverages conformalized quantile regression to output calibrated predictive intervals, providing statistically robust per-star rotation uncertainty metrics. This rolling-window inference reveals that 31,953 stars (21.5$\%$) exhibit bimodal rotational signals. By incorporating APOGEE $v \sin i$ measurements, we demonstrate that for distinct (non-harmonic) bimodals, the longer mode represents the true rotation, exposing a systematic failure mode wherein classical single-pass periodograms lock onto shorter aliases. Filtering by our calibrated confidence intervals yields a highly reliable subset of 119,428 stars.  The catalog resolves various rotation-related phenomena: the metallicity dependence of
rotation at fixed stellar mass, pointing on the role of metallicity in magnetic braking processes; tracing
equatorial velocity and specific angular momentum directly across the Kraft
break; recovery of empirical gyrochronology sequences and identification
of hierarchical triple candidates among the synchronized-binary population.
\emph{The Maunder} provides reliable rotation periods for the largest main-sequence population in \textit{Kepler}, allowing for population-level studies of rotation-based phenomena.
\end{abstract}

%% file: sections/intro.tex
\section{Introduction}\label{sec:intro}
Rotation is a fundamental stellar property that connects various stellar processes through the evolution of a star. During their main-sequence lifetime, stars with a convective envelope lose angular momentum through magnetized wind in a process called \emph{magnetic braking} \citep{Schatzman_1962, Weber_1967, Skumanich_1972}, linking rotation period to stellar age, magnetic activity, and internal structure. The inference of stellar age from its rotation rate is usually called \emph{gyrochronology} and is an active area of research \citep{barnes_rotational_2003, barnes_ages_2007, mamajek_improved_2008, angus_calibrating_2015, angus_toward_2019, bouma_empirical_2023, bouma_ages_2024, lu_age_2024, Van-Lane_2025_chronoflow} that reveals non-trivial relationships between rotation period, stellar type, and age.\\
Rotation period is also important for exoplanet research; Stellar rotation period is induced by spot modulation on the surface of the star and can hinder real exoplanet signals or even be detected as false planets \citep{Robertson2014_activity_as_planet, Nava_2020_exoplanet_limitation, Jeffers_2022_carmenes_exoplanets}. Moreover, the rotation period, when combined with radius and $v\sin(i)$  measurements, can be used to derive stellar inclination, which is important for spin-orbit alignment studies in exoplanets \citep{Mazeh2015_inclination, Walkowicz2013_exoplanets}.\\
Rotation period is also an important factor in close binary systems, where tidal torque transfers angular momentum from the binary to the star, in a process that circularizes the orbit and synchronizes the stellar and orbital periods \citep{Zahn_1977, Hut1981, Ogilvie_2014, Mazeh_2008_tides}. As such, the rotation period is an important part of the study of tidal evolution and binary populations \citep{Bashi2023_tides, Simonian_2019, Lurie_2017}.\\
The rotation information is manifested in a stellar light curve due to magnetic activity on the surface of the star; active regions with strong and vertical magnetic fields suppress the convection of heat flux from the core to the surface of the star, leaving a darker area called \emph{spot} \citep{solanski_2000_spots, Berdyugina2005StarspotsAK}. A group of spots can have a lifetime of up to hundreds of days \citep{Namekata2019_spots_lifetime}. When a spot's lifetime is long compared to the rotation rate of the star, we get periodic variations in a light curve that measure the surface-integrated flux from the star. Those variations are the basis for period inference from light curves. However, the variations are often not fully periodic but only quasi-periodic and depend on the spot lifetime, which is not always sufficiently long, and on the star's magnetic activity, which might change over the observation period (e.g., \cite{Reinhold2017_activity}, \cite{Giles2017_spotlife}, \cite{Basri2022_spotlife}). In addition, stellar light curves usually suffer from various noise sources which stem from both astrophysical and instrumental origins \citep{Gilliland2011_kepler_noise, pont2006_planet_noise, Dalba2017_crowding}. A long line of work dealt with how to optimally extract rotational signal from light curves, using classical signal analysis methods, such as the Lomb-Scargle periodogram \citep{Lomb1976, Scargle1982}, autocorrelation function \citep{McQuillan2013_acf}, and wavelet transform \citep{Reinhold_2022_GPS}.\\
The \emph{Kepler} mission \citep{Borucki2010_kepler_mission} produced light curves for approximately $200{,}000$ stars, spanning $4$ years of observation with $30$ minutes cadence measurements. Its combination of long baseline, high photometric precision, and continuous coverage makes it uniquely suited for precise rotation measurement. Indeed, the implementation of classical methods led to the first catalogs of rotation periods in the Kepler field \citep{McQuillan2014_acf_catalog, Santos2021_catalog, Reinhold2023_gps_catalog}, which enabled population analysis of rotation-related phenomena. However, the complex manifestation of periodicity in the light curves and the rigidity of model-based signal processing methods lead to inconsistency between the different methods (see, for example, \citealt{Lu2022_meta_analysis}), resulting in limited scalability.\\
Machine learning models suggest a data-driven approach that might provide more flexibility and might be able to generalize beyond standard signal processing. In recent years, there has been a growing interest in the applications of machine learning to stellar light curves. For example, machine learning has been used to denoise light curves \citep{Morvan2022_denoise}, detect exoplanets \citep{Garvin2024_planets, Zucker2018_planets_dl}, classify variable stars \citep{Kang2023_variable_stars}, and predict stellar parameters \citep{Pan2024_astroconf, Zuo2026_falco}. \\
As part of this broader phenomenon, several studies have developed machine learning models specifically for period detection. One of the earliest examples is \cite{Blancato2022_cnn}, which used a convolutional neural network (CNN) to predict stellar period. While pioneering, they used a simple CNN and trained their model only on labels from a specific catalog (\cite{McQuillan2014_acf_catalog}), which limits flexibility. In a slightly different approach, \cite{Breton2021_rf} and \cite{Gomes2024_xgb} used classical machine learning algorithms as vetting and interpolation algorithms between classical methods. \cite{Claytor2022_dl_butter} and \cite{Claytor2024_tess} used a CNN trained on synthetic light curves with realistic noise from the \emph{TESS} survey \citep{ricker2015_tess}. The high noise and limited observation time in TESS made their model applicable only for a small subset of TESS samples and not transferable to Kepler samples. \cite{Kamai2025_lightpred} used a self-supervised pre-training step; training that does not rely on labels and extracts representations based on similarity metrics. However, the fine-tuning stage still relied on simulated light curves, and the residual gap between simulation and real data introduced inconsistencies with previous catalogs. This limitation directly motivates the present approach, in which supervised labels are drawn entirely from cross-catalog consensus rather than simulation. In more recent works, \cite{Zuo2026_falco} and \cite{Ding2026_starclr} used fully self-supervised models, but those models were not trained or fine-tuned to predict periods.\\
There are various challenges when applying data-driven models for period predictions. First, the labels used during training define an upper bound on the accuracy of the model. A model that was trained on a specific catalog can be only as good as the catalog, and would learn the systematics and errors that come with it. As mentioned above, inconsistencies between catalogs suggest that those biases are not negligible and that a smart interpolator might be able to generalize better. On the other hand, using simulated light curves removes ambiguity about labels but introduces a bigger problem - the gap between simulation and real data. Since the physics of spot modulation and their manifestation in light curves is not yet fully understood, and since any simulation must make concrete assumptions, this gap is inevitable. A further challenge concerns the self-supervised objective itself. Such models learn from similarity or reconstruction alone, aiming to produce a representation useful for downstream fine-tuning (self-supervised learning is therefore often called \emph{representation learning}). The two dominant families have distinct failure modes. Reconstruction-based methods, which predict masked or future parts of the input \citep{He2021_mae, Devlin2018_bert, Radford2018_gpt}, are sensitive to noise; joint-embedding (energy-based) methods, which organize the latent space so that similar samples lie close together \citep{Chen2020_simclr, Chen2020_simsiam, Caron2021_dino, Benton2017_deepCCA, Bardes2021_vicreg}, can instead collapse to constant or low-rank representations. For a detailed comparison see \cite{VanAssel2025_pred_vs_align}. In the astrophysical context, \cite{Zuo2026_falco} is reconstruction-based, whereas \cite{Kamai2025_lightpred} and \cite{Ding2026_starclr} are joint-embedding models.\\
The choice between families is not merely technical but tied to the physical task. Different stellar parameters imprint on the light curve at different time scales: surface gravity ($\log g$) traces granulation, which varies on scales much shorter than the rotation period, whereas magnetic modulation evolves on much longer ones. A reconstruction-based model trained to predict a short window should therefore recover $\log g$ much better than the rotation period, which depends on long-time-scale structure. This consideration directly motivates our model, which uses an energy-based objective and multi-scale input design.\\
A final challenge is interpretability. Classical methods such as the ACF and LS periodogram provide mechanistic reliability diagnostics, for instance the ratio between the dominant peak and the background, that let users assess and filter individual predictions. Machine-learning models rarely expose such direct diagnostics. Providing a calibrated, per-star measure of confidence is therefore essential if a data-driven catalog is to support population studies, which is one of the central aims of this work. \\
To summarize, using machine learning for stellar parameter inference introduces various challenges related to the credibility of labels, the gap between simulation and real data, uncertainty estimation, and the differing time scales of distinct physical phenomena.\\
In this work, we address those challenges and introduce a new machine learning model for stellar period prediction, which we dub \emph{The Maunder}\footnote{Named for Annie Scott Dill Maunder (n\'ee Russell,
1868--1947), an Irish astronomer whose contribution to the sunspot ``butterfly diagram'',
published in \citet{Maunder1904_butterfly} under her husband's name, went
uncredited in her lifetime \citep{Dalla2016_annie_maunder}.}. The Maunder is trained with multiple self-supervised and supervised objectives simultaneously. The supervised labels are drawn from samples that are consistent between at least two existing catalogs. The combination of the two objectives reduces the ambiguity and single-catalog systematics, while keeping the self-supervised objective on samples without labels, which ensures the model is trained on all types of samples and not only the high-quality ones. Since many classical methods work in the frequency domain, we split the light curve into frequency-based inputs and time-based inputs. The frequency-based inputs are the ACF and LS of the light curve. The time-based inputs are the raw light curve under different normalizations and smoothing windows, which emphasize the different time scales. In addition, our model predicts not a single period but a set of period quantiles, trained with a quantile-regression objective and calibrated through conformal prediction \citep{Romano2019_cqr}, so that each star is assigned a predictive interval with calibrated coverage whose width grows for ambiguous or noisy light curves. We further aggregate predictions across multiple segments of each light curve and report the scatter between segments, providing a complementary, empirical probe of prediction stability that is sensitive to non-stationary spot evolution and to the model's robustness to the observed window. We treat these two quantities as complementary reliability probes. Rather than imposing predefined quality cuts, we publish the full catalog together with both uncertainty estimates, enabling users to define selection functions appropriate to their scientific task.\\
The paper is organized as follows: in section \ref{sec:data}, we present the dataset and pre-processing procedures. In section \ref{sec:model}, we present our model and training pipeline. In section \ref{sec:results}, we present our main results, in section \ref{sec:catalog}, we present the catalogs, and in section \ref{sec:conclusions}, we conclude, discuss limitations, and future directions.

%% file: sections/data.tex
\section{Data}\label{sec:data}
We use long-cadence light curves from the \emph{Kepler} mission Data Release~25
\citep{kepler_dr25}, corrected for instrumental systematics with the PDC-MAP
pipeline \citep{smith2012_pdc-map}. The data are available at
MAST\footnote{\url{https://doi.org/10.17909/T9488N}}. We work with the PDCSAP
flux on its native ${\approx}30$-minute cadence ($48$ samples per day), using quarters $3-16$, and
reconstruct time on a uniform grid, with missing cadences set to zero.

\subsection{Rotation Dataset}\label{sec:dataset}
To create a reliable rotation dataset, we filter the entire Kepler dataset using the
color-magnitude classification \texttt{flag\_CMD} of
\citet{Godoy-Rivera2025}. Our dataset consists of all stars
not classified as subgiants or giants, $148{,}746$ stars. We
separate the populations because surface rotation in evolved stars occupies a
different period regime and is usually not characterized by spot modulations.

Supervised labels are assigned only to stars with a \emph{consensus} rotation
period, which we define as agreement to within $20\%$ between at least two of the
reference catalogs of \citet{McQuillan2014_acf_catalog},
\citet{Santos2021_catalog}, \citet{Reinhold2023_gps_catalog}, and \cite{Kamai2025_lightpred}. The final period label is the average period over all consensus catalogs. This yields
$41{,}650$ labeled main-sequence stars. Requiring cross-catalog consensus suppresses the single-catalog systematics as discussed in Section~\ref{sec:intro}, at the cost of a smaller labeled set; the remaining stars participate through the self-supervised objective alone
(Section~\ref{sec:model}). The regression target is $\log_{10}(P_{\mathrm{rot}})$, a choice motivated by the wide dynamic range of rotation periods.

From each light curve we draw windows of $450$ days ($21{,}600$ long-cadence
samples). The window length balances two competing requirements: it must span
several cycles of the longest rotation periods of interest, while remaining
short enough that two non-identical windows can be drawn from a single
${\sim}4$-year baseline. The latter is essential because our self-supervised
objective requires two distinct views of the same star
(Section~\ref{sec:model}). During training we therefore draw two random
$450$-day windows per star, and zero-pad missing observations.
Each window is converted into a six-channel input that exposes the rotation signal in complementary domains. All channels derive from a Savitzky--Golay
smoothed version of the flux (window ${\approx}0.5$~d, first order), which
suppresses sub-rotational scatter while preserving the spot-modulation envelope.

\begin{figure}[t]
    \centering
    \includegraphics[width=0.5\linewidth]{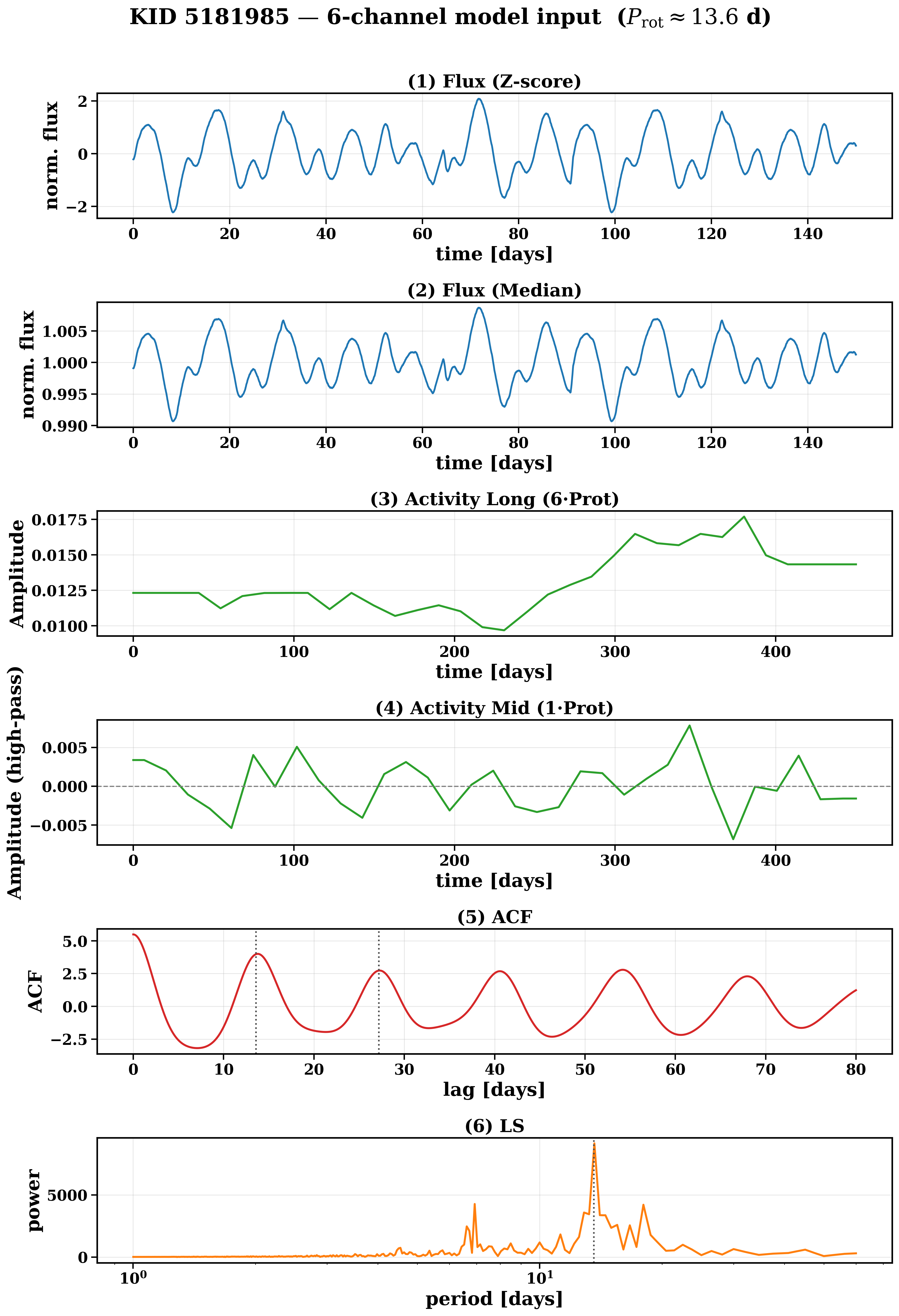}
    \caption{Example of the 6 input channels for Kepler $5181985$. Each panel shows a specific channel input. For more details, see Section~\ref{sec:dataset} }\label{fig:channels}
\end{figure}

Two \emph{flux} channels capture the modulation shape, as a z-scored series, and
its fractional amplitude, as a median-normalised series. Z-score is a popular normalization for machine learning models as it standardizes the input in a way that highlights morphological changes. However, stellar light curves are usually median-normalized because it preserves relative amplitudes. We decided to keep both normalization as separate channels. In addition, we add two \emph{activity-proxy} channels that measure the variability amplitude on
rotation-related timescales. Each proxy is computed as the difference between the $5th$ and $95th$ percentiles over different time windows. The percentile difference is median normalized and log-transformed. Percentile difference was shown by \cite{Reinhold2017_activity} as an activity proxy. Importantly, the different time windows correspond to different activity time scales (e.g., \citealt{Maunder1904_butterfly}, and \citealt{Rieger1984_rieger}) and require an initial guess for the rotation period. We call the initial guess period $P_{\mathrm{ref}}$, and elaborate on the process of acquiring $P_{\mathrm{ref}}$ in~\ref{sec:data_scaffold}. The long time-scale was derived with a window of $6P_{\mathrm{ref}}$, and the mid time-scale was derived with a window of one $P_{\mathrm{ref}}$. For the mid time-scale, we further applied a high-pass filter by subtracting the average over 10 time-scales. This removes the long-time-scale signature.
The \emph{autocorrelation function} (ACF) and \emph{Lomb--Scargle Periodogram} (LS) channels
provide the classical frequency-domain rotation diagnostics
\citep{McQuillan2013_acf, Lomb1976, Scargle1982}, computed on the z-scored flux;
The full input consists of a 6-channel light curve, with 4 time-domain channels and 2 frequency-domain channels, and is shown in Figure \ref{fig:channels}.

\subsection{Activity-proxy window scaffold}\label{sec:data_scaffold}
The window and stride of the two activity-proxy channels are scaled by a
per-star reference period $P_{\mathrm{ref}}$, so that the proxies track the
expected rotation timescale rather than a fixed duration.
We obtain $P_{\mathrm{ref}}$ with the following \emph{scaffolding} procedure: we first train a four-channel version of the same model, using only the channels that require no period scaffold
(the two flux channels, ACF, and LS), and use its median period prediction,
frozen per star, to size the activity-proxy windows of the full six-channel
model. We emphasize that $P_{\mathrm{ref}}$ is a flux-derived estimate from the
model's own earlier stage, enters \emph{only} through the geometry of the
activity-proxy windows, is held fixed across all crops of a given star, and is
never used as a supervised target; the regression labels are completely independent from $P_{\mathrm{ref}}$ as they are the consensus periods of Section~\ref{sec:dataset}. We additionally expose the
$80\%$ prediction-interval width of the four-channel stage as a single scalar
input to the prediction head, where it acts as a per-star reliability hint for
the scaffold period. In Section~\ref{sec:results} we verify that the
six-channel model does not simply reproduce $P_{\mathrm{ref}}$, and even improve the scaffold predictions when they are wrong.

%% file: sections/model.tex
\section{Model}\label{sec:model}
The model used to predict the periods is a machine learning model trained with both self-supervised and supervised objectives. The self-supervised objective is applied to each sample in the dataset, and the supervised objective is added to the $41{,}650$ samples with consensus period label. Contrastive self-supervised models are trained to maximize similarity between different views of the same objects. The two views are usually passed through a shared encoder, and the model is optimized based on a similarity metric between the two features. Our model follows a similar architecture with some important changes. The two views in our case are two random windows drawn from the same Kepler observation, each one with $6$ channels, as described in~\ref{sec:data}. The views are processed through a shared encoder that splits each view into time-domain channels (the two flux normalizations and the two activity proxies) and frequency-domain channels (ACF and LS). The time-domain channels are processed through an AstroConformer module \citep{Pan2024_astroconf}, which combines transformers and CNN blocks. Importantly, this module uses RoPE positional encoding \citep{Su2021_rope} to keep the time-domain information. During this module, the sequence length is reduced by a CNN filter with a stride of approximately one day ($49$ points), which results in a sequence of $440$ tokens, each one with a dimension of $1024$. The frequency-domain channels are processed with a simple CNN encoder with $6$ layers, resulting in a sequence of $337$ tokens, each one with a dimension of $1024$. The two sequences are concatenated together to form a sequence of $777$ tokens and sent to a conformer mixer \citep{Gulati2020_conformer}, with RoPE positional encoding on the tokens. \\
The final tokens from each view are sent into a DualFormer module. DualFormer was presented in \cite{Kamai2025_desa} as a module to combine multimodal views of stellar light curves and stellar spectra. It consists of both cross- and self-attention blocks, mean pooling over the token dimension, addition of meta information, and two separate linear projection layers. The original paper also introduced a new loss objective, called \emph{duality loss}, that serves as a different similarity metric. This objective defines similarity based on the expectation value of the pooled
features with respect to the projection layer, and it was shown to outperform other objectives on downstream astrophysical tasks.
We use the same DualFormer module with two changes. First, in \citet{Kamai2025_desa}
the two projection layers are constrained to be the transpose of one another, motivated
by the light-curve/spectrum duality; since we use only light-curve views, we relax this
constraint and treat them as two independent, asymmetric linear maps $A$ and $B$.
Second, we add a variance term that explicitly regularises the variance of the bilinear interaction, constraining its spread across the batch to prevent the
interaction from collapsing to a constant.

Concretely, let $\mathbf{e}^{(1)}_i,\mathbf{e}^{(2)}_i$ be the pooled embeddings of the
two views of star $i$ in a batch of size $B$, and $\mathbf{p}^{(1)}_i=A\,\mathbf{e}^{(1)}_i$,
$\mathbf{p}^{(2)}_i=B\,\mathbf{e}^{(2)}_i$ their projections. Each view is a
scalar bilinear interaction,
\begin{equation}
a_i = \langle \mathbf{p}^{(1)}_i, \mathbf{e}^{(1)}_i\rangle, \qquad
b_i = \langle \mathbf{p}^{(2)}_i, \mathbf{e}^{(2)}_i\rangle .
\end{equation}
The duality (matching) term ties the two views together, and a variance hinge keeps
each interaction from collapsing to a constant,
\begin{align}
\mathcal{L}_{\rm match} &= \frac{1}{B}\sum_{i=1}^{B}(a_i-b_i)^2, \\
\mathcal{L}_{\rm var} &= \mathrm{ReLU}\!\left(1-\sigma_a\right) + \mathrm{ReLU}\!\left(1-\sigma_b\right), \\
\sigma_a &= \sqrt{\mathrm{Var}_i(a_i)+\epsilon}
\end{align}
with $\sigma_b$ defined analogously and $\epsilon=10^{-4}$, giving the duality objective
\begin{equation}
\mathcal{L}_{\rm dual} = \mathcal{L}_{\rm match} + \lambda_{\rm var}\,\mathcal{L}_{\rm var},
\qquad \lambda_{\rm var}=1 .
\end{equation}
Collapse of the projection vectors themselves is prevented separately by a covariance
term $\mathcal{L}_{\rm cov}$: with centred projections
$\tilde{\mathbf{p}}^{(v)}=\mathbf{p}^{(v)}-\tfrac{1}{B}\sum_i\mathbf{p}^{(v)}_i$ and the
within- and cross-view covariances
$C_1=\tilde{\mathbf{p}}^{(1)\top}\tilde{\mathbf{p}}^{(1)}/(B-1)$,
$C_2=\tilde{\mathbf{p}}^{(2)\top}\tilde{\mathbf{p}}^{(2)}/(B-1)$,
$C_{12}=\tilde{\mathbf{p}}^{(1)\top}\tilde{\mathbf{p}}^{(2)}/(B-1)$,
\begin{equation}
\mathcal{L}_{\rm cov} = \frac{1}{D}\left[
\sum_{i\neq j}(C_1)_{ij}^2 + \sum_{i\neq j}(C_2)_{ij}^2 + \sum_{i\neq j}(C_{12})_{ij}^2 \right],
\end{equation}
where $D$ is the projection dimension; the within-view terms are the standard VICReg
covariance and $C_{12}$ additionally decorrelates feature dimensions across the two views. In Table \ref{tab:ssl_ablation}, we show that our objective performs better than standard VICReg objective.

\begin{figure*}\centering
\resizebox{0.7\linewidth}{!}{\input{sections/figs/activity_v2_simsiam_architecture}}
\caption{Diagram of our model. A light curve is split into two windows, processed by a shared-weights encoder. The two views are then sent into a DualFormer module and to prediction heads. }\label{fig:model}
\end{figure*}
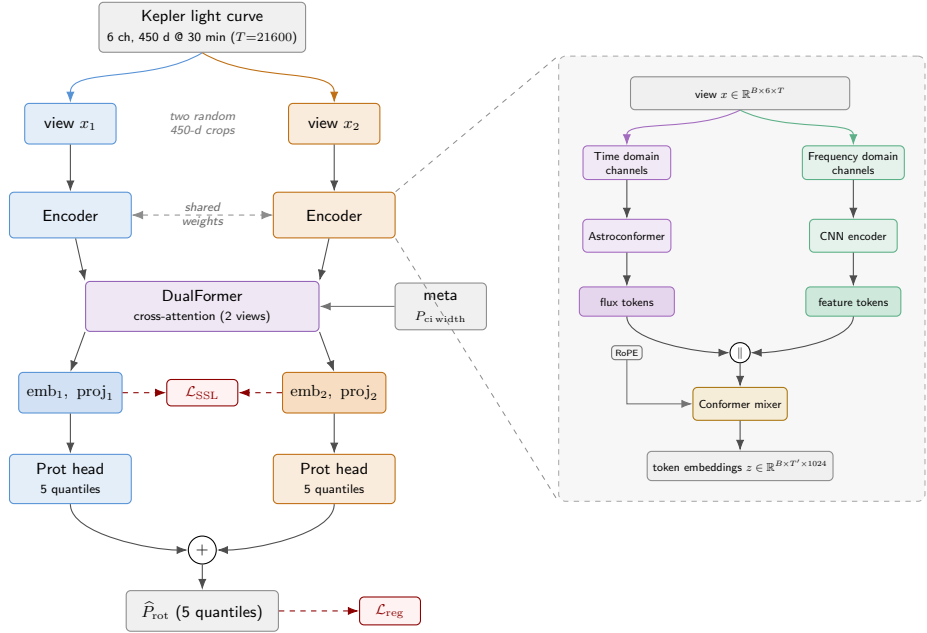

For supervision, we add a small MLP head to each branch output. The head predicts $5$
quantiles of $\log_{10}P_{\rm rot}$, and the final prediction is the average of the two
branches. The use of quantiles rather than a point estimate is what enables uncertainty
estimation. The supervised term is the multi-quantile pinball loss
\begin{align}
\mathcal{L}_{\rm reg} &= \frac{1}{|\mathcal{Q}|}\sum_{q\in\mathcal{Q}}
\rho_q\!\left(y-\hat{y}_q\right), \\
\rho_q(u) &= \max\!\big(q\,u,\,(q-1)\,u\big),
\end{align}
over the quantile levels $\mathcal{Q}=\{0.1,0.25,0.5,0.75,0.9\}$ with $y=\log_{10}P_{\rm rot}$,
averaged over the labelled stars; the resulting intervals are calibrated post-hoc by split
conformalized quantile regression \citep{Romano2019_cqr}, which guarantees marginal coverage. The calibration was done on the validation set, so the test set is unaffected.
The total training objective is
\begin{equation}
\mathcal{L} = \tfrac{1}{2}\,\mathcal{L}_{\rm reg}
            + \tfrac{1}{4}\,\mathcal{L}_{\rm dual}
            + \tfrac{1}{4}\,\mathcal{L}_{\rm cov}.
\end{equation}
Figure~\ref{fig:model} shows a schematic diagram of our model. 

%% file: sections/figs/activity_v2_simsiam_architecture.tex
%
%

\definecolor{viewA}{RGB}{30,110,200}   
\definecolor{viewB}{RGB}{225,130,20}   
\definecolor{dfh}{RGB}{120,60,165}     
\definecolor{cnnc}{RGB}{38,150,95}     
\definecolor{astc}{RGB}{140,70,175}    
\definecolor{mix}{RGB}{200,140,20}     

\begin{tikzpicture}[
  font=\sffamily\small,
  >={Latex[length=2.2mm]},
  blk/.style ={rounded corners=2.5pt, draw, semithick, align=center,
               inner sep=4.5pt, minimum width=24mm, minimum height=9mm},
  inp/.style ={blk, fill=black!6,    draw=black!45},
  VA/.style  ={blk, fill=viewA!12,   draw=viewA!75},
  VB/.style  ={blk, fill=viewB!15,   draw=viewB!85!black},
  D/.style   ={blk, fill=dfh!12,     draw=dfh!80},
  lat/.style ={blk, minimum width=18mm, minimum height=8mm},
  op/.style  ={circle, draw, semithick, inner sep=1pt, minimum size=5.5mm, fill=white},
  a/.style   ={->, semithick, black!70},
  loss/.style={rounded corners=2.5pt, draw=red!55!black, semithick, fill=red!5,
               align=center, inner sep=4pt, font=\sffamily\footnotesize,
               text=red!55!black},
  dl/.style  ={dashed, ->, red!55!black, semithick},
  sub/.style ={font=\sffamily\scriptsize\itshape, text=black!55, align=center},
  C/.style   ={blk, fill=cnnc!12,  draw=cnnc!75},
  A/.style   ={blk, fill=astc!12,  draw=astc!80},
  M/.style   ={blk, fill=mix!16,   draw=mix!85!black},
  tok/.style ={blk, minimum width=26mm, minimum height=8mm},
  rope/.style={rounded corners=2pt, draw=black!55, semithick, fill=black!4,
               inner sep=3pt, font=\sffamily\scriptsize, align=center},
]


\node[inp, minimum width=40mm] (raw) at (0,0.4) {Kepler light curve\\\scriptsize 6 ch, 450 d @ 30 min ($T{=}21600$)};
\node[VA, lat] (v1) at (-2.6,-1.5) {view $x_1$};
\node[VB, lat] (v2) at ( 2.6,-1.5) {view $x_2$};
\node[sub] at (0,-1.5) {two random\\450-d crops};

\node[VA] (encA) at (-2.6,-3.3) {Encoder};
\node[VB] (encB) at ( 2.6,-3.3) {Encoder};
\node[sub, text=black!60] (sh) at (0,-3.3) {shared\\weights};
\draw[<->, dashed, black!45] (encA.east) -- (encB.west);

\node[D, minimum width=46mm] (df) at (0,-5.1) {DualFormer\\\scriptsize cross-attention (2 views)};
\node[inp, lat] (meta) at (4.7,-5.1) {meta\\\scriptsize $P_{\mathrm{ci\,width}}$};

\node[VA, lat, fill=viewA!20] (e1) at (-2.6,-6.8) {$\mathrm{emb}_1,\ \mathrm{proj}_1$};
\node[VB, lat, fill=viewB!26] (e2) at ( 2.6,-6.8) {$\mathrm{emb}_2,\ \mathrm{proj}_2$};
\node[loss, minimum width=14mm] (ssl) at (0,-6.8) {$\mathcal{L}_{\mathrm{SSL}}$};

\node[VA] (h1) at (-2.6,-8.5) {Prot head\\\scriptsize 5 quantiles};
\node[VB] (h2) at ( 2.6,-8.5) {Prot head\\\scriptsize 5 quantiles};
\node[op] (avg) at (0,-9.9) {$+$};
\node[inp, lat, minimum width=30mm] (pred) at (0,-11.1) {$\widehat{P}_{\mathrm{rot}}$ (5 quantiles)};
\node[loss, minimum width=12mm] (cqr) at (3.7,-11.1) {$\mathcal{L}_{\mathrm{reg}}$};

\draw[a, viewA!75]       (raw.south) .. controls +(-1.0,-0.6) and +(0,0.7) .. (v1.north);
\draw[a, viewB!85!black] (raw.south) .. controls +( 1.0,-0.6) and +(0,0.7) .. (v2.north);
\draw[a] (v1) -- (encA);   \draw[a] (v2) -- (encB);
\draw[a] (encA) -- (df.north west);  \draw[a] (encB) -- (df.north east);
\draw[a, black!55] (meta) -- (df.east);
\draw[a] (df.south west) -- (e1.north);  \draw[a] (df.south east) -- (e2.north);
\draw[a] (e1) -- (h1);   \draw[a] (e2) -- (h2);
\draw[a] (h1.south) .. controls +(0,-0.6) and +(-0.7,0) .. (avg.west);
\draw[a] (h2.south) .. controls +(0,-0.6) and +( 0.7,0) .. (avg.east);
\draw[a] (avg) -- (pred);
\draw[dl] (e1.east) -- (ssl.west);
\draw[dl] (e2.west) -- (ssl.east);
\draw[dl] (pred.east) -- (cqr.west);

\begin{scope}[shift={(10.6,-0.9)}, scale=0.72, transform shape]

  \node[inp, minimum width=60mm] (x) at (0,0)
       {view $x\in\mathbb{R}^{B\times 6\times T}$};

  \node[A] (xflux) at (-3.1,-1.9) {Time domain\\channels};
  \node[C] (xfeat) at ( 3.1,-1.9) {Frequency domain\\channels};

  \node[A] (astro) at (-3.1,-3.9) {Astroconformer};
  \node[C] (cnn)   at ( 3.1,-3.9) {CNN encoder};

  \node[A, tok, fill=astc!22] (zA) at (-3.1,-5.7) {flux tokens};
  \node[C, tok, fill=cnnc!22] (zC) at ( 3.1,-5.7) {feature tokens};

  \node[op]   (cat)  at (0,-7.1) {$\Vert$};
  \node[rope] (rope) at (-3.1,-7.1) {RoPE};
  \node[M]    (mixer) at (0,-8.5) {Conformer mixer};
  \node[inp, tok, minimum width=48mm] (z) at (0,-10.2)
       {token embeddings $z\in\mathbb{R}^{B\times T'\times 1024}$};

  \draw[a, astc!80] (x.south) .. controls +(-1.2,-0.6) and +(0,0.7) .. (xflux.north);
  \draw[a, cnnc!75] (x.south) .. controls +( 1.2,-0.6) and +(0,0.7) .. (xfeat.north);
  \draw[a] (xflux) -- (astro);   \draw[a] (xfeat) -- (cnn);
  \draw[a] (astro) -- (zA);      \draw[a] (cnn)   -- (zC);
  \draw[a] (zA.south) .. controls +(0,-0.7) and +(-0.7,0) .. (cat.west);
  \draw[a] (zC.south) .. controls +(0,-0.7) and +( 0.7,0) .. (cat.east);
  \draw[a, black!55] (rope.south) |- (mixer.west);
  \draw[a] (cat) -- (mixer);
  \draw[a] (mixer) -- (z);

  \begin{scope}[on background layer]
    \node[draw=black!35, dashed, rounded corners=4pt, fill=black!3,
          fit=(x)(xflux)(xfeat)(astro)(cnn)(zA)(zC)(cat)(rope)(mixer)(z),
          inner sep=4mm] (encbox) {};
  \end{scope}
\end{scope}

\draw[dashed, black!45, thin] (encB.north east) -- (encbox.north west);
\draw[dashed, black!45, thin] (encB.south east) -- (encbox.south west);

\end{tikzpicture}

%% file: sections/results.tex
\section{Results}\label{sec:results}
We trained our model on the main-sequence dataset as described in Section~\ref{sec:data}, after a split into $80\%,10\%,10\%$ for train, validation, and test sets. Because the model was trained to predict periods from two 450-day windows, and most Kepler observations span longer baselines, we generated final predictions using a rolling-window procedure. We evaluated the model on consecutive windows with a stride of 90 days and aggregated the per-window predictions. In this inference pipeline, the two input views are identical. For the large majority of stars the per-window periods form a single cluster, and the adopted period is the median across windows; a minority instead split into two well-separated clusters, and for these \emph{bimodal} stars the adopted period follows the convention defined in Section~\ref{sec:bimodal}. To capture the model's sensitivity to local observational noise, we report the standard deviation of $\log_{10}(P_{\rm{rot}})$ across the windowed predictions, $\sigma_{\mathrm{win}}$, as an additional metric of inherent predictive scatter.\\

\begin{figure*}
    \centering
    \includegraphics[width=0.5\linewidth]{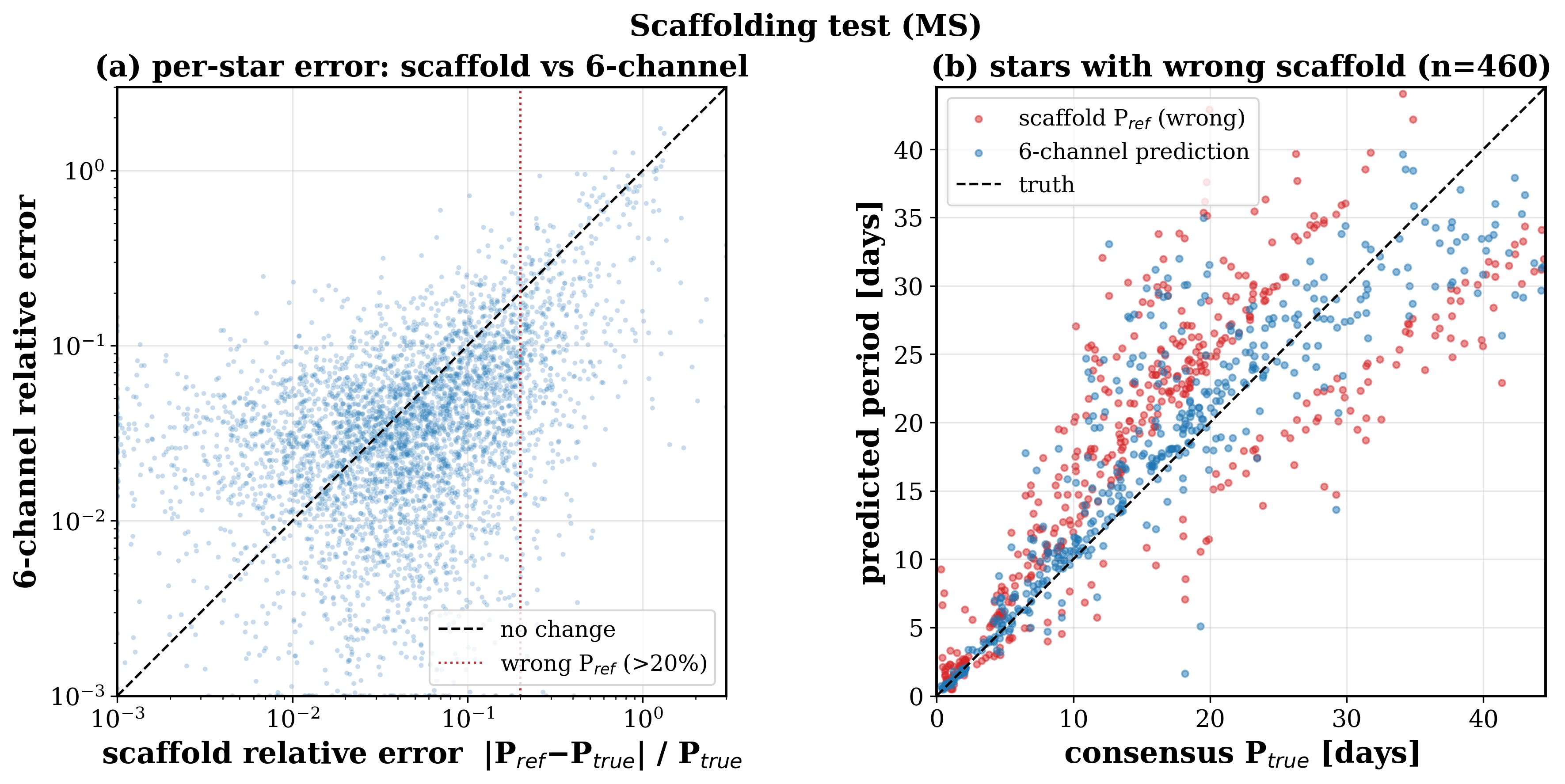}
    \caption{Testing scaffolding period, $P_{\mathrm{ref}}$, against the full model predictions. The right panel shows $460$ samples for the held-out test set, where $|P_{\mathrm{ref}}-P_{\mathrm{true}}|\geq 20\%$. The x-axis is $P_{\mathrm{true}}$, and the y-axis shows $P_{\mathrm{ref}}$ in red and predictions of the full model, which uses $P_{\mathrm{ref}}$ as a scaffold, in blue. It can be seen that the blue points have lower error. The left panel shows the scaffold error as a function of the full model error for the entire test set. The red dashed line marks $20\%$ error of the scaffold period, so all the points to the left of this line are in the right panel. }\label{fig:scaffold_test}
\end{figure*}

First, we test the scaffolding procedure. As mentioned in \ref{sec:data_scaffold}, the activity proxies use a scaffold period, $p_{\mathrm{ref}}$, which comes from a prior inference of the same model without activity proxies. We call this prior model \emph{scaffold model}. While $p_{\mathrm{ref}}$ is not given as a label but only sets a timescale for activity proxies, we want to test how it affect the results. Figure \ref{fig:scaffold_test} shows a comparison between the full model and the scaffold model on the test set. The right panel shows the true period ($P_{\mathrm{true}}$) vs. scaffold period ($P_{\mathrm{ref}}$,  red) and predicted period of the full model ($P_{\mathrm{pred}}$, blue), for 460 samples with large scaffold error ($|P_{\mathrm{ref}}-P_{\mathrm{true}}| > 0.2 P_{\mathrm{true}}$). It can be seen that the blue points are much closer to the $x=y$ line, which implies that the full model improves the wrong scaffolds. Indeed, in $86\%$ of the cases, the error of the full model is lower compared to the error of the scaffold model. The left panel shows the scaffold error vs. the full model error for the full test set. The dashed red line marks the criteria for the right panel -  $|P_{\mathrm{ref}}-P_{\mathrm{true}}| > 0.2 P_{\mathrm{true}}$. We see that in the region of small errors ($1\%-10\%$), there are samples where the predicted period increases the error. However, as our labels were constructed with $20\%$ consensus between catalogs, this might reflect natural ambiguity in the labels. \\

\subsection{Performance evaluation}\label{sec:performance}
\begin{figure*}
    \centering
    \includegraphics[width=0.5\linewidth]{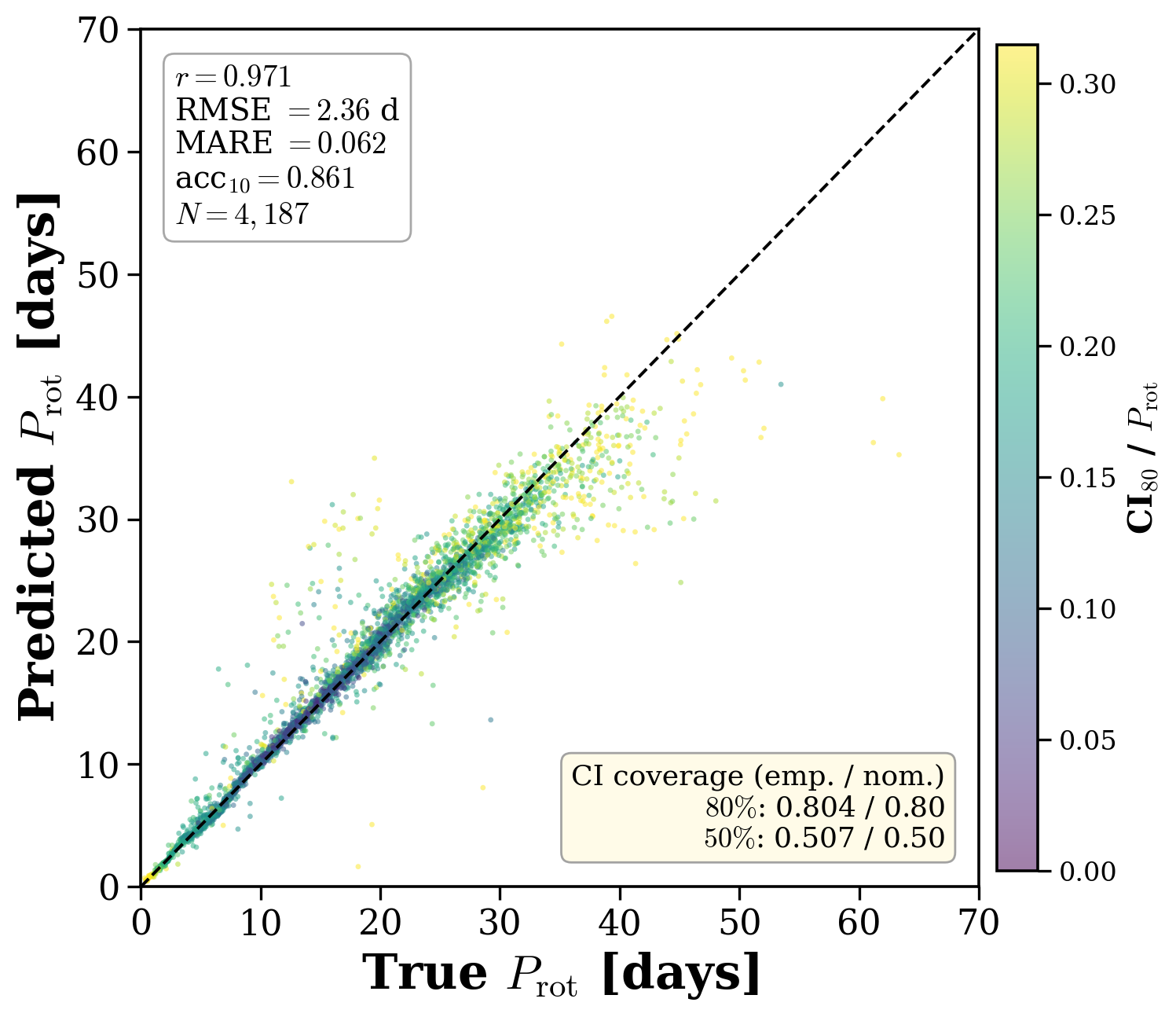}
    \caption{true period vs. predicted period (median quantile) for our model}\label{fig:true_vs_pred}
\end{figure*}

We test the full model predictions on the test set. Figure \ref{fig:true_vs_pred} shows the true period vs. the predicted period, evaluated on a held-out test set, for our model. The colors represents the normalized confidence interval predicted by the model - difference between $90^{th}$ and $10^{th}$ quantiles, divided by the predicted period. It can be seen that the agreement is very good. Our model reaches RMSE of $2.36$ days, better than \cite{Blancato2022_cnn} ($5.2$ days) and \cite{Kamai2025_desa} ($2.61$ days). However, it is important to note that both \cite{Blancato2022_cnn} and \cite{Kamai2025_desa} didn't use the same training labels. Specifically, they didn't use consensus labels. Another important observation is that the confidence intervals (CI) are correctly calibrated - we see the empirical coverage match the expected coverage. The median absolute error is $0.46$~d. For comparison, the mutual scatter of the
input catalogs on their agreeing subsets implies a single-catalog error of
$\sim\!1.1$~d, so the model's typical prediction is more precise than any
individual catalog it learns from. The squared error is dominated by a small number of harmonic failures. Of the 4,187 test stars, 28 ($0.67\%$) differ from their label by close to a factor of two and together account for $20.9\%$ of the squared error; 24 are doubled and 4 halved. Only 5 of the 28 are flagged bimodal (see ~\ref{sec:bimodal}), and the CI$_{80}/P_{\rm rot} <
0.4$ cut retains 26, so the reliability metrics of Section~\ref{sec:ci} do not
isolate them.\\
Another evaluation of the model is by inspecting its latent space. Reducing the dimensionality of our $1024$ latent dimensions using PCA and UMAP \citep{McInnes2018_umap}, reveals physical structure, with clear gradients in period value and period coherence (we elaborate more on coherence in~\ref{sec:low_coherence}), and a clear separation between seemingly similar populations, like the separation between young fast rotators and synchronized binaries (see \cite{Kamai2025_too_fast}, and \cite{Murphy2026_fast_rotators}). UMAP examples are shown in Appendix~\ref{sec:appendix} (Figure~\ref{fig:umap_2d} and Figure~\ref{fig:umap_3d}).

\subsection{low-coherence rotators}\label{sec:low_coherence}
\begin{figure*}
    \centering
    \includegraphics[width=0.7\linewidth]{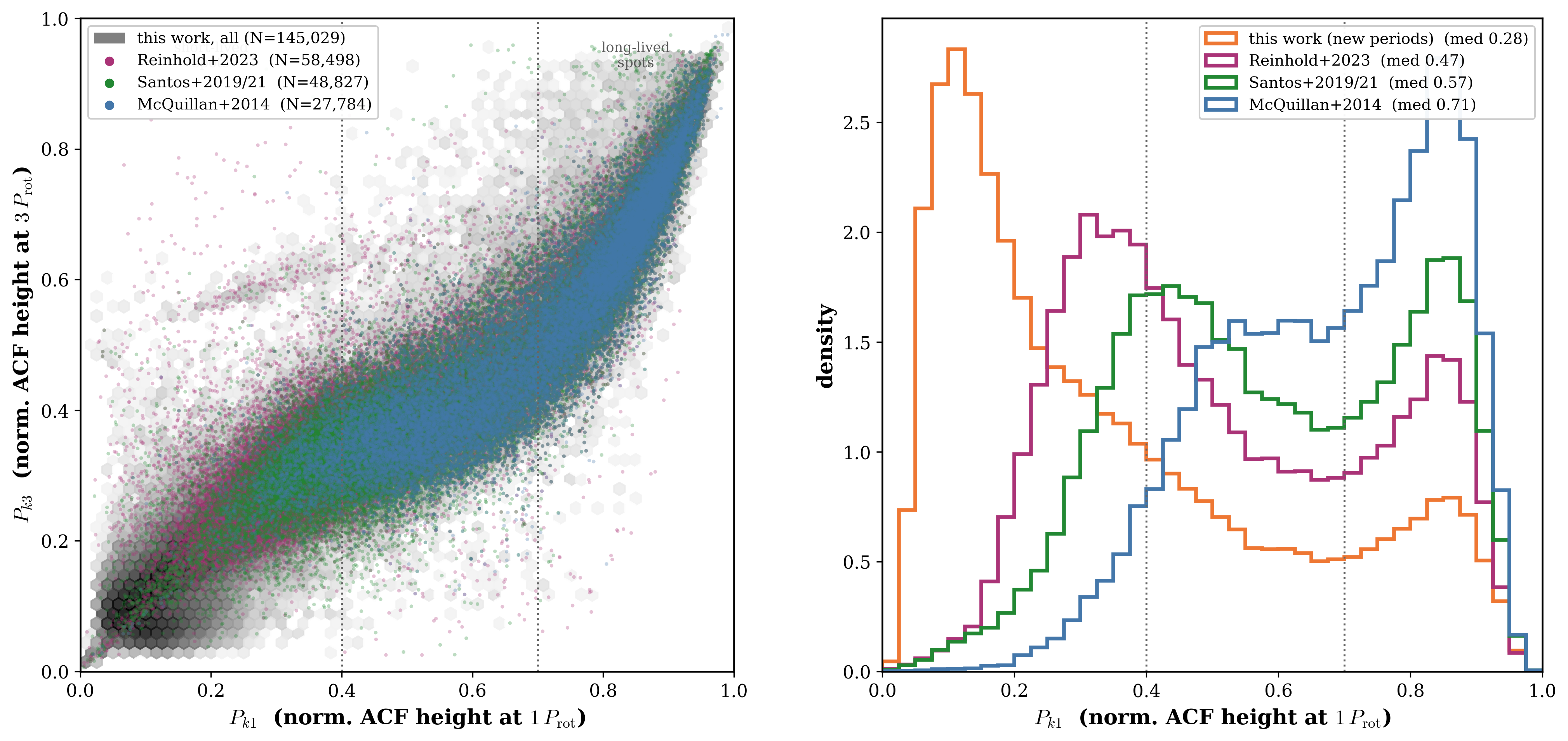}
    \caption{Analysis of coherence using the first and third ACF normalized peaks. The left panel is a peak-to-peak plot (PPP), which shows the relationship between peaks for different catalogs, including our new catalog. \cite{Basri2022_spotlife} used such a plot to analyze spots' lifetimes. The right panel shows the distribution of the first peak height, which we refer to as a \emph{coherence factor}. It can be seen that newer catalogs tend to use less coherent populations.}\label{fig:ppp}
\end{figure*}

\begin{figure*}
    \centering   \includegraphics[width=0.7\linewidth]{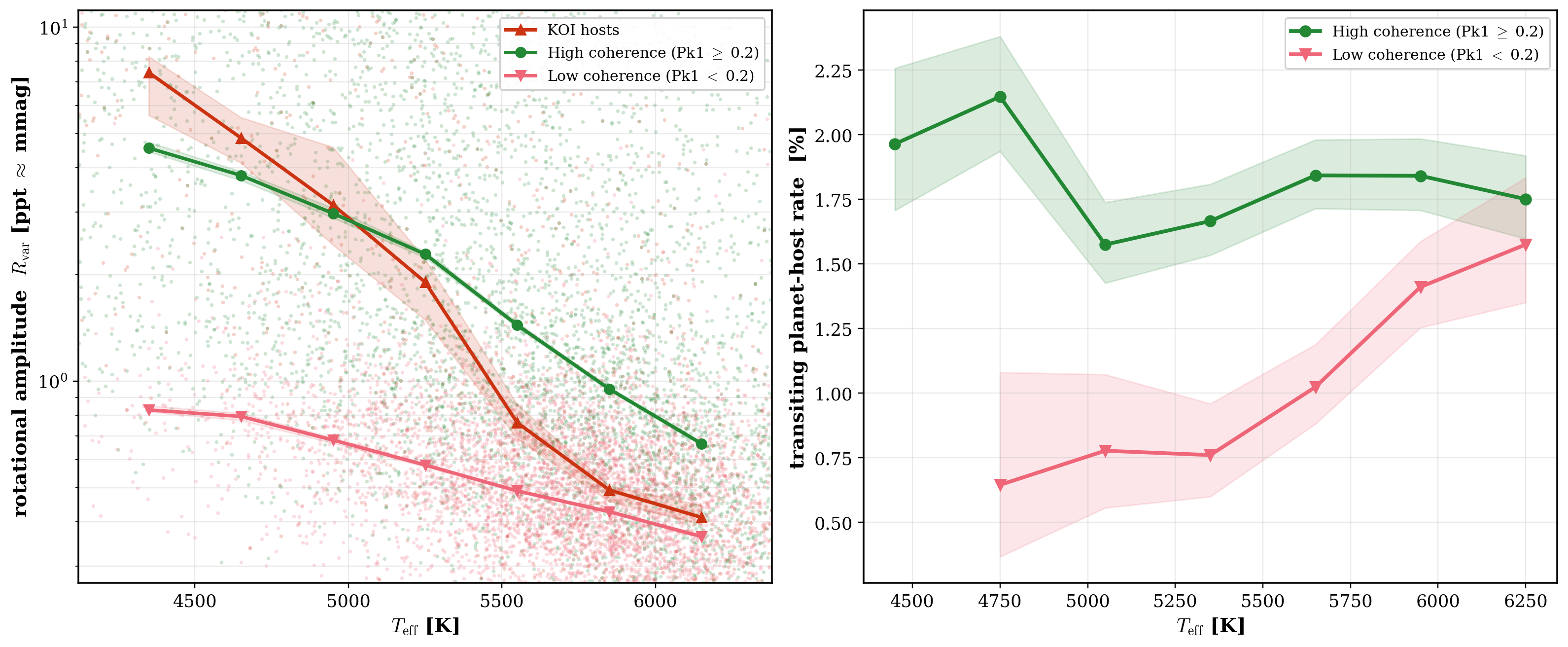}
    \caption{Left panel - photometric amplitude vs. $T_{\rm{eff}}$. The different colors represent different populations - high coherence unimodal stars (normalized ACF peak $\geq 0.2$), low coherence unimodal stars (normalized ACF peak $< 0.2$), and planet host stars. The solid curves are the median over temperature bins. Error bars are estimated from bootstrapping - resample each bin with replacement and report one $\sigma$ interval of the medians. Right panel - planet host fraction as a function of $T_{\rm{eff}}$ for the same populations. Error bars are Poisson errors}\label{fig:inclination}
\end{figure*}

Increasing the size of the period catalog introduces populations with more challenging periodicity. One measure of how much a star has a coherent periodicity was presented by \cite{Basri2022_spotlife}. This is simply the height of the strongest ACF peak after normalization. \cite{Basri2022_spotlife} used this parameter to identify spot lifetime, and we will come back to this type of analysis in \ref{sec:bimodal}, but for now we treat it as a general \emph{coherence parameter}. Figure~\ref{fig:ppp} demonstrates, on the right panel, the different coherence distributions in different catalogs. The trend is clear - newer and bigger catalogs include less coherent samples. The left panel shows the same trend in a peak-to-peak (ppp) plot, comparing the first and third ACF peaks. \cite{Basri2022_spotlife} used this kind of plot to identify regions of spot lifetimes: the upper right corner represents long-lived spots, and the lower left corner is short-lived spots. Both panels demonstrate that in our dataset there is a large sample of low-coherence stars. We note that $p_{k1}$ is not independent of the photometric amplitude ($R_{\rm{var}}$): across our sample the two are strongly correlated (Spearman $\rho=0.79$), so the
low-$p_{k1}$ population is largely a low-amplitude one. Coherence should
therefore be read as an amplitude-related quantity rather than an independent property of the light curve.\\ 
The last point connects coherence with inclination. When the star has low inclination, fewer spots are visible, and variability is weaker. This effect was demonstrated in \cite{Mazeh2015_inclination} using analysis of the photometric amplitude of transit planet hosts compared to the general population. Transit planets are biased to be detected at high orbital inclination, and because of spin-orbit alignment, this bias propagates to the inclination of their hosts. Interestingly, spin-orbit alignment breaks around the Kraft break \citep{Albrecht2022_spin_orbit}, so we expect planet hosts to have high inclinations only below the Kraft break. This is exactly the trend in photometric amplitude, found in \cite{Mazeh2015_inclination} (Figure 1 in their paper). We therefore expect that low-coherence samples would be identified as low-inclination, using a similar analysis to the one presented in \cite{Mazeh2015_inclination}. While we reproduce their analysis on the general population (Appendix~\ref{sec:appendix}, Figure~\ref{fig:inclination_mazeh_reproduce}), separation into high and low coherence populations, using a threshold of $0.2$, produces the expected result (left panel in Figure~\ref{fig:inclination})---we see that low coherence samples have a much lower photometric amplitude, compared to high coherence and planet hosts. We also see that around the Kraft break, where spin-orbit is approximately isotropic (see \cite{Albrecht2022_spin_orbit}), planet hosts sit between high and low coherence populations. This supports the correlation between high-low coherence and high-low inclinations, at population level. The right panel of Figure~\ref{fig:inclination} complements the picture by looking at the planet host fraction in high and low coherence samples. We see that for cool stars, the planet-host fraction is approximately a factor of two higher for high-coherence stars than for low-coherence and bimodal stars, and this gap closes towards the Kraft break.

\subsection{Bimodal rotators}\label{sec:bimodal}

Because the final period is inferred on rolling windows (Section~\ref{sec:results}), a star whose light curve contains two well-separated rotational signals, or whose signal is ambiguous between a true period and an alias, appears not as a single period with scatter, but as two distinct clusters of per-window periods. KID~892376 (Fig.~\ref{fig:bimodal_lc}) is a clear example: an early segment dominated by a $\approx1.5$~d modulation,\footnote{For KID~892376 the short mode appears as $1.4$~d (the $2$-means cluster mean) and $1.5$~d (the aggregate over windows); the two differ because the aggregate is a window median while the mode value is a within-cluster mean.} of the kind recovered by \cite{McQuillan2014_acf_catalog}, and a late segment showing a $\approx14$~d modulation, of the kind recovered by the \cite{Reinhold2023_gps_catalog} ACF. That each mode is independently reproduced by a different classical catalog already indicates the bimodality is a property of the light curve rather than an artifact of our model.

We flag such stars automatically. For every star with at least four windows we work in $\log_{10}P$ and partition the per-window periods with a one-dimensional $2$-means split (scanning all sorted cut points). A star is labelled \emph{bimodal} when the two modes are separated by at least $0.20$~dex (a factor $1.6\times$), the minority mode holds at least $20\%$ of the windows, and the split is clean: either the separation exceeds three times the within-mode scatter, or the silhouette of the two-way assignment exceeds $0.60$. Splits consistent with a $2{:}1$ ratio ($|\Delta\log_{10}P-\log_{10}2|<0.06$~dex) are labelled harmonic aliases; the remainder are treated as distinct bimodal.

Of the $148{,}746$ main-sequence stars, $31{,}953$ ($21.5\%$) are bimodal, of which $25{,}357$ ($79\%$) show distinct periods and $6{,}596$ ($21\%$) show the $2{:}1$ harmonic behaviour. The ratio between the two modes has a median ratio of $3.2$, with $64\%$ of bimodals separated by more than $2.5\times$, so the majority are not simple harmonics. Because a typical bimodal star spans two modes, its $80\%$ interval is wide by construction, and the $\mathrm{CI}_{80}/P_{\rm rot}<0.4$ cut (Section~\ref{sec:ci}) removes bimodals more aggressively than unimodals ($53.2\%$ vs.\ $87.7\%$ retained).

\paragraph{Which mode is the rotation.} Photometry alone cannot break this degeneracy, so we bring in an independent spectroscopic indicator, the projected rotation velocity $v\sin i$. For a candidate period $P$ and radius $R$ the implied inclination is $\sin i = \frac{v \sin i }{v_\mathrm{eq}} = \frac{v\sin i \cdot P}{2\pi R}$, which must not exceed unity and, over an isotropically-oriented population, follows $\cos i \sim \mathcal{U}(0,1)$. We cross-match the distinct bimodals to APOGEE~DR17 $v\sin i$ \citep{APOGEE_DR17_2022} and \citet{Berger2020_kepler_prop} radii, and build a hierarchical forward model \citep[following][]{MasudaWinn2020_hbm} that places the macroturbulent broadening, the ASPCAP measurement-error floor, and the ~1.5 km/s detection truncation inside the likelihood. Its free parameters are calibrated on a control sample of $N\simeq2{,}790$ unimodal (unambiguous-period) stars, which the calibrated model reproduces to KS $D=0.12$ (Fig.~\ref{fig:vsini_hbm_control},  Appendix~\ref{sec:appendix}); because APOGEE $v\sin i$ is inflated near its floor, only the comparison against this control is meaningful, not absolute $\sin i$.

We then ask which mode, taken as the rotation period, reproduces the observed $v\sin i$ distribution. For the $N=390$ distinct bimodals with $v\sin i$, adopting the long mode matches the observations as well as the control does (KS $D=0.10$), while adopting the short mode predicts far too many rapid rotators ($D=0.48$)(Fig.~\ref{fig:vsini_hbm}). The $2{:}1$ harmonic splits, by contrast, are \emph{not} resolved by $v\sin i$: the two hypotheses differ by only $|\Delta D|=0.05$, smaller than the $D=0.12$ control-calibration residual, and the per-star odds are near unity ($54\%$ favor long, see Figure~\ref{fig:vsini_hbm_harmonic}, Appendix~\ref{sec:appendix}). We also did not find any significant difference in the correlation to other physical properties between the two harmonic modes, which might point to the true period among them. We conclude that harmonic bimodals are ambiguous.

We therefore adopt the long mode as the rotation period for distinct bimodals, and leave the harmonic splits at their aggregate period with an ambiguity flag. The short mode is thus not a second rotation rate but an alias, plausibly related to a short time-scale phenomenon. 

The adopted-period convention has a direct consequence: for a distinct bimodal, the median-over-windows aggregate is the majority mode, which is frequently the shorter alias rather than the rotation. The published catalog therefore carries, for every star, the bimodal flag, the split type, and both mode periods, so that users can revert the adoption or weight by it; the hard adopted period is the long mode for distinct splits.

\begin{figure*}
    \centering
    \includegraphics[width=0.5\linewidth]{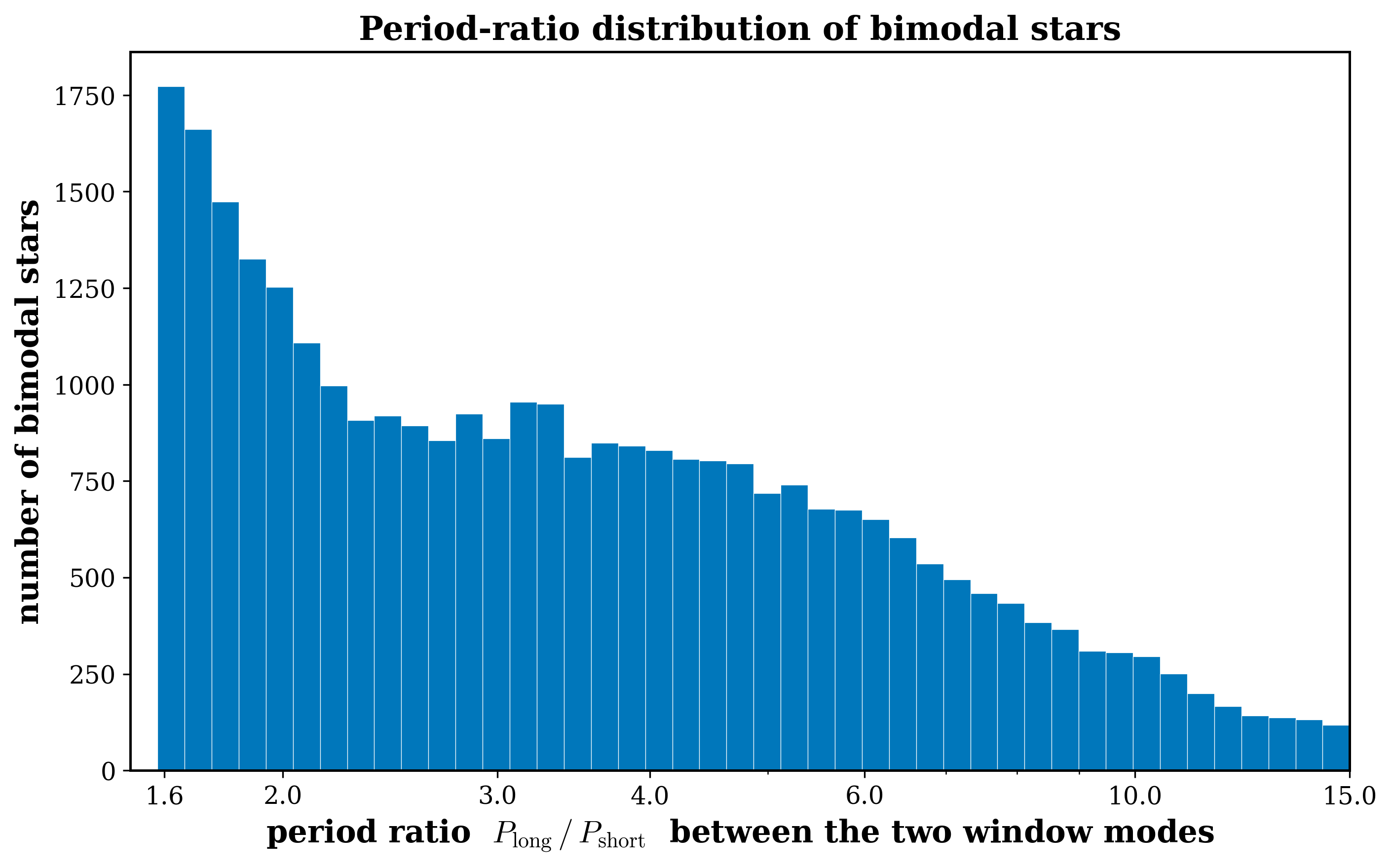}
    \caption {Distribution of period ratios in bimodal stars.}\label{fig:bimodal_p_ratio}
\end{figure*}

\paragraph{Physical origin of the low mode}
We want to investigate the origin of the bimodal population.
We begin by excluding harmonic aliases.
Figure~\ref{fig:bimodal_p_ratio} shows the distribution of mode ratios: a smooth
continuum from $1.6$ out to $15$, with no excess at $2{:}1$ or $3{:}1$ beyond the harmonic class defined above. Discrete harmonic aliasing therefore cannot account for the distinct bimodals.
Next, we exclude contaminants, starting with eclipse contamination. Cross-matching against the Kepler Eclipsing Binary
Catalog \citep{Kirk2016_keplerEB}, detached eclipsing binaries are strongly depleted among bimodals relative to unimodal stars ($0.19\%$ vs.\ $1.05\%$,
Fisher odds ratio $0.18$), that is, in the opposite sense to what eclipse
injection would produce. We note that this comparison is conservative rather
than clean: an eclipse is a strictly repeating signal, so eclipsing systems
are preferentially classified unimodal by the coherence criterion itself.
The direction of that bias can only deepen the apparent deficit, so the
exclusion of eclipse contamination is robust to it.
Next, aperture contamination is excluded. \texttt{PDCSAP} removes only the median flux of a neighbouring star, so a \emph{variable} contaminant survives
the correction, and because Kepler rolled by $90\degr$ each quarter such a
contaminant would enter and leave the aperture in quarter-length blocks. We therefore compared the \texttt{CROWDSAP} keyword (the fraction of aperture flux
belonging to the target) across quarters $3$--$16$ for $700$ distinct
bimodals and $700$ unimodal stars matched in $T_{\rm eff}$ and magnitude. The two classes are indistinguishable in both contamination level (median
\texttt{CROWDSAP} $0.9992$ vs.\ $0.9993$; Mann--Whitney $p=0.97$) and, more importantly, in quarter-to-quarter variability (spread
$0.0032$ vs.\ $0.0033$; $p=0.71$). The short mode is not a neighbouring
star's signal entering through a varying aperture. This test is sensitive to flux from resolved neighbours within the aperture and would not detect a
companion blended inside the point-spread function core.\\
A possible hypothesis to the cause of bimodality is that the spot lifetime is short relative to the rotation period: when active regions survive many rotations, the period is locked (unimodal), and when they live only a rotation or two, the signal
decoheres, and the per-window period destabilizes. We test this with the spot-lifetime estimator of \cite{Basri2022_spotlife}, which maps the coherence parameter to a spot-group lifetime, using $5th$ order polynomial, and expressed either in days ($L_{\rm day}$) or rotations ($L_{\rm rot}=L_{\rm day}/P_{\rm rot}$).
Having identified the long mode as the true rotation (above), we compute $L_{\rm day}$ and $L_{\rm rot}$ for the entire population, unimodal and bimodal, using each star's true period.
In the left panel of Figure~\ref{fig:L_day}, we bin the bimodal fraction against $L_{\rm day}$: it rises monotonically from $\sim2\%$ at
$L_{\rm day}\gtrsim100$~d to $\sim55\%$ at $L_{\rm day}\approx2$~d, so stars whose spots survive only a few days are far more likely to be flagged bimodal. We show in Figure~\ref{fig:L_rot} (Appendix~\ref{sec:appendix}) that the same trend holds in the
dimensionless lifetime $L_{\rm rot}=\tau_{\rm spot}/P_{\rm rot}$, so the effect is not an
artifact of the period.
Next, we check the $T_{\rm eff}$ dependence. The right panel in Figure~\ref{fig:L_day} shows the median $L_{\rm{day}}$, aggregated over $T_{\rm{eff}}$ bins. We see single-period stars trace the expected relation discussed in \cite{Basri2022_spotlife} and \cite{Giles2017_spotlife}---
median $L_{\rm day}$ rises from $\sim20$~d at $6600$~K to $\sim160$~d toward the cool
end, recovering the longer spot lifetimes of cool dwarfs.
Bimodals, however, sit well below at every temperature, and are much less temperature dependent. While Figure~\ref{fig:L_day} might seem convincing, it hides an important caveat - the spot-lifetime estimator of \cite{Basri2022_spotlife} was calibrated on the \cite{McQuillan2014_acf_catalog} samples alone, namely, only the most coherent stars (see Figure~\ref{fig:ppp}). Using it for bimodals, which have low coherence, is a strong extrapolation that might break its applicability. This might explain the low $T_{\rm{eff}}$ sensitivity in Figure~\ref{fig:L_day} as a consequence of the uncalibrated polynomial rather than physical phenomena. Therefore, without recalibration of the \cite{Basri2022_spotlife} expression, we cannot conclusively conclude that short spot lifetime is the origin of the low mode. Such calibration is left for future work.  

\begin{figure*}
    \centering
    \includegraphics[width=\linewidth]{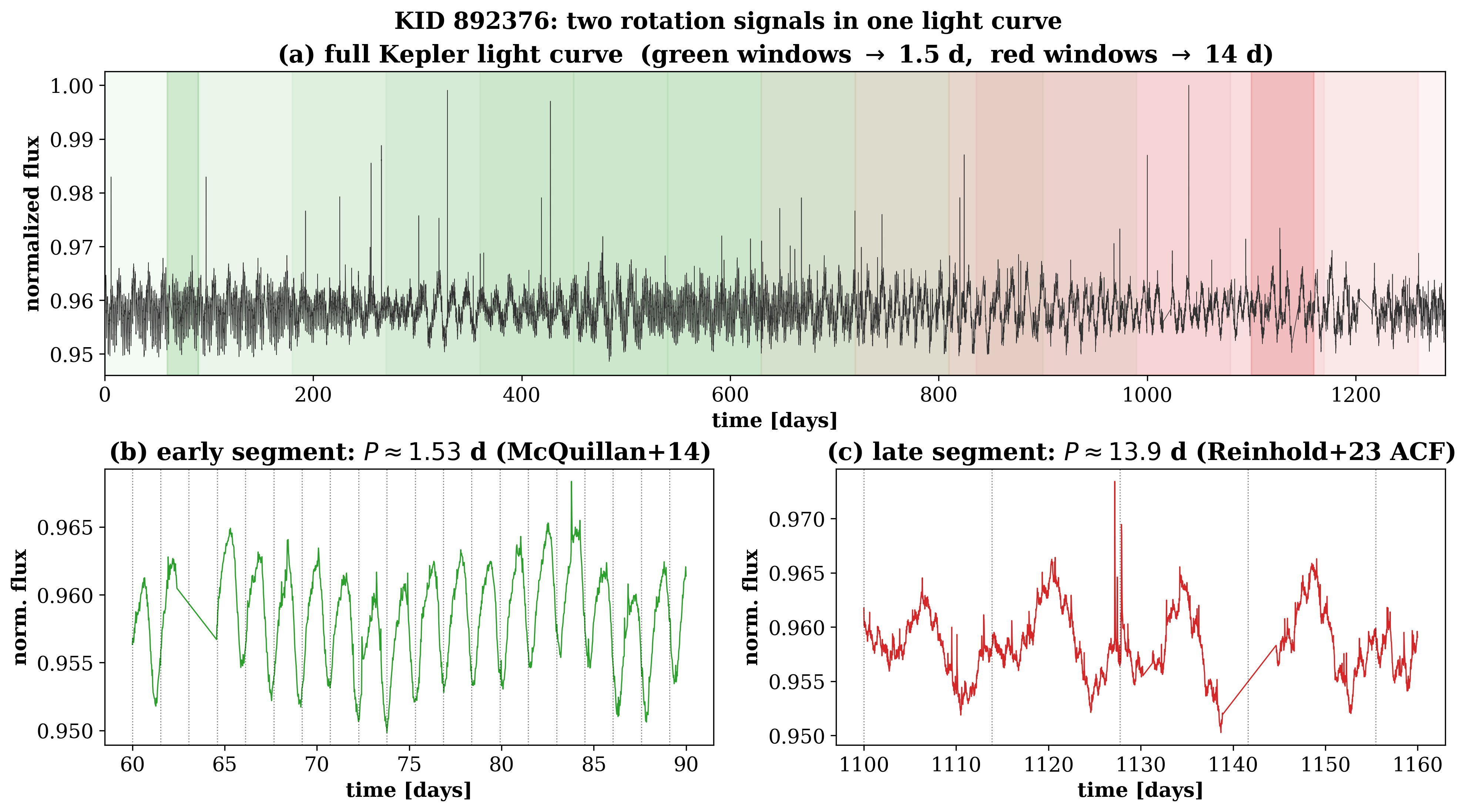}
    \caption{Upper panel - full light curve of KID $892376$. The shaded colors represent the different window predictions of our model (green - lower period). Lower panel - zoom in of two segments with very different periods. Gray dotted lines mark the predictions of \cite{McQuillan2014_acf_catalog} and \cite{Reinhold2023_gps_catalog} that fit the two segments. }\label{fig:bimodal_lc}
\end{figure*}

\begin{figure*}
    \centering
    \includegraphics[width=0.75\linewidth]{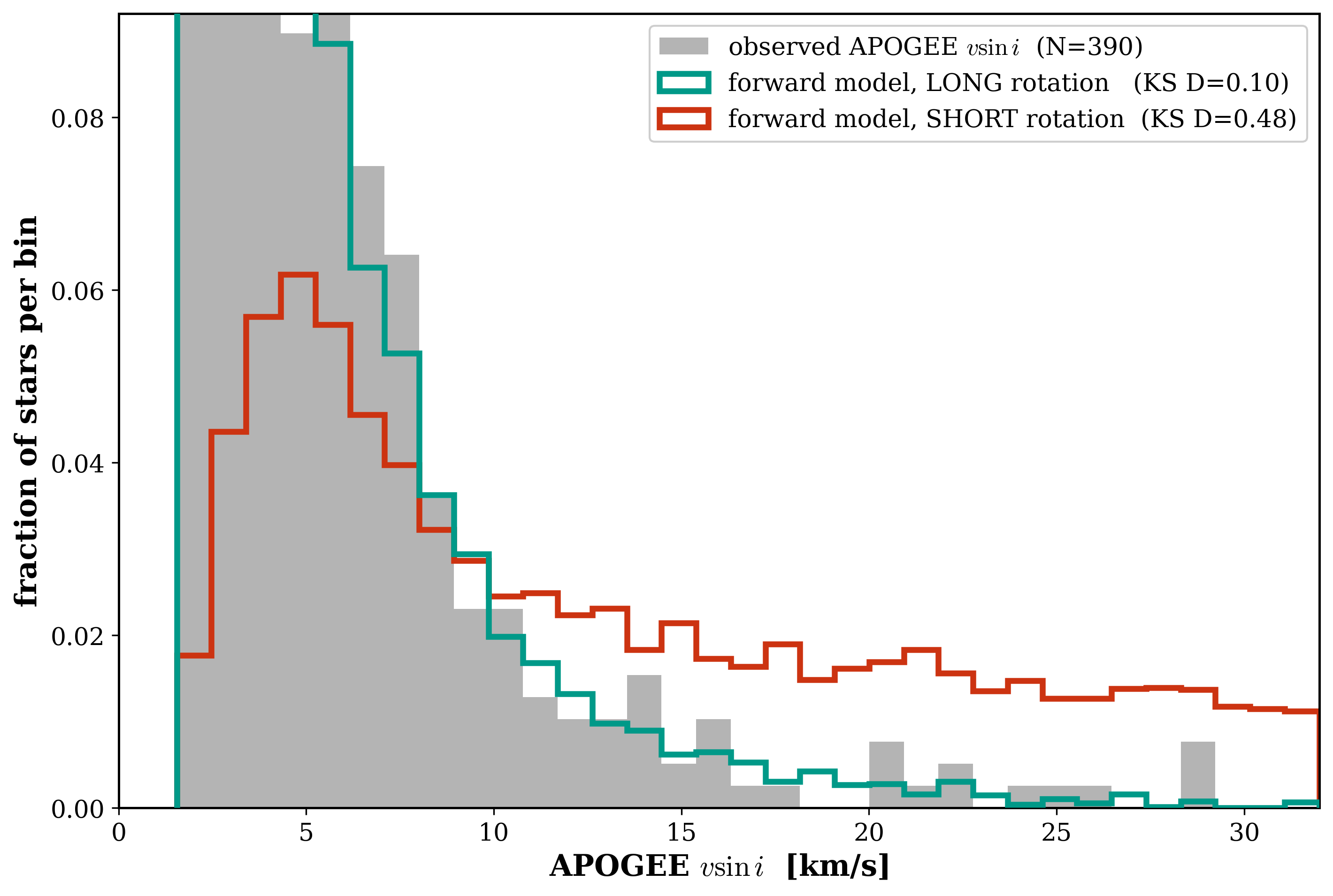}
    \caption{Adjudicating the true rotation mode with APOGEE $v\sin i$ ($N=390$ distinct bimodals). Grey: observed $v\sin i$. Coloured: the hierarchical forward-model prediction \citep[following][]{MasudaWinn2020_hbm} when the \emph{long} (teal) or \emph{short} (red) mode is taken as the rotation period. Adopting the long mode reproduces the observed distribution as well as the control does (KS $D=0.10$), while the short mode predicts far too many rapid rotators ($D=0.48$). The forward-model floor/macroturbulence parameters are calibrated on a $N\simeq2{,}790$ unambiguous-period control (KS $D=0.12$).}\label{fig:vsini_hbm}
\end{figure*}

\begin{figure*}
    \centering
    \includegraphics[width=0.7\linewidth]{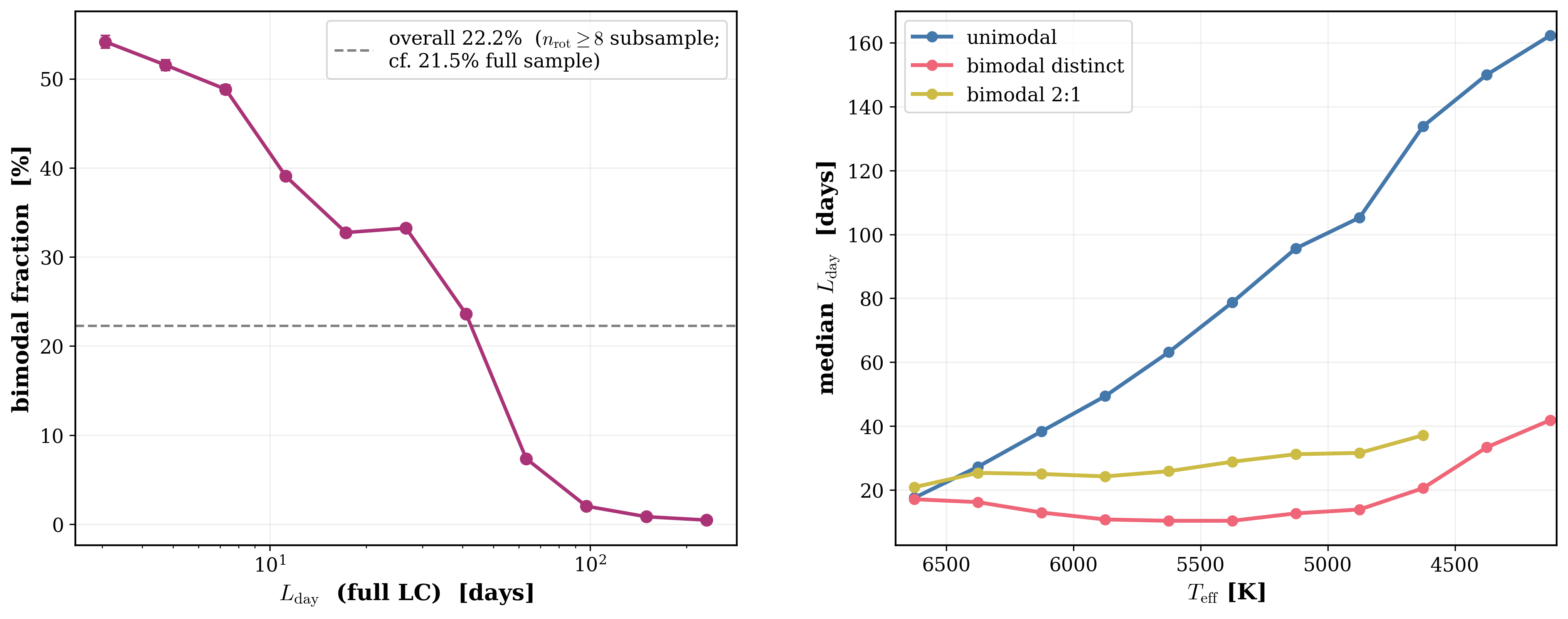}
    \caption{Spot-group lifetime and bimodality. Left - The bimodal fraction rises monotonically as the spot-group lifetime in days, $L_{\rm day}$ (from the first normalized-ACF peak height following \citealt{Basri2022_spotlife}), approaches unity; dashed line marks the overall fraction. Right - Median $L_{\rm day}$ as a function of $T_{\rm eff}$: single-period (unimodal) stars rise monotonically for cool stars, while bimodal stars stay relatively flat.}\label{fig:L_day}
\end{figure*}

\subsection{Confidence intervals as a proxy to model uncertainty}\label{sec:ci}
\begin{figure*}
    \centering
    \includegraphics[width=0.7\linewidth]{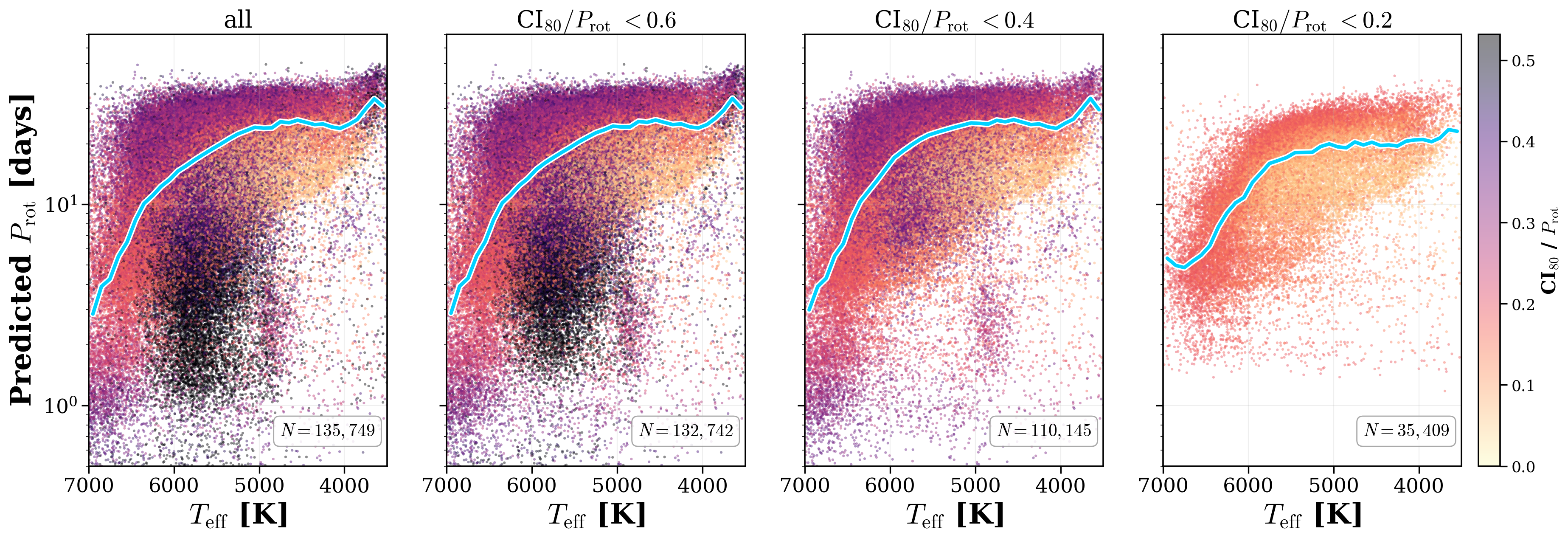}
    \caption{$T_{\mathrm{eff}}$ vs. $P_{\mathrm{rot}}$ , colored by normalized $80\%$ CI ($\frac{P_{0.9}-P_{0.1}}{P_{\mathrm{rot}}}$). Each panel uses a different upper bound on the normalized $80\%$ CI. The number of points that satisfy the bound is written in each panel. The light blue line is the median $P_{\mathrm{rot}}$ over bins of $T_{\mathrm{eff}}$. }\label{fig:teff_prot_ci80}
\end{figure*}

\begin{figure*}
    \centering
    \includegraphics[width=0.7\linewidth]{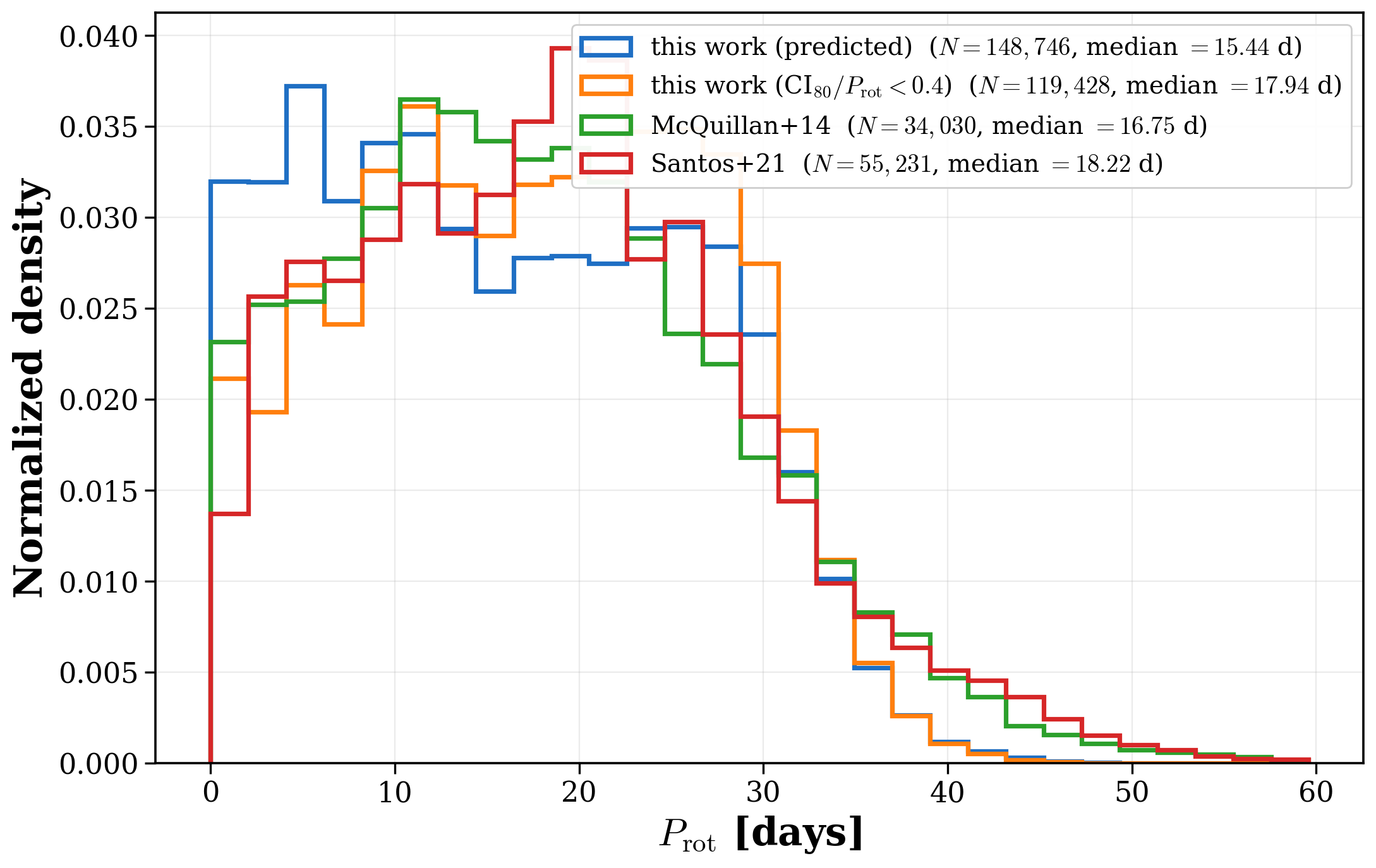}
    \caption{period distribution of \cite{McQuillan2014_acf_catalog} (green), \cite{Santos2021_catalog} (red) and our main sequence predictions. The blue histogram shows the full main sequence dataset, and the orange histogram shows a filtered dataset, keeping only samples with $\text{CI}_{80}/P_{\mathrm{rot}}<0.4$   }\label{fig:prot_dist_compare}
\end{figure*}

As mentioned in \ref{sec:model}, our model predicts not only a point estimate of the period, but also quantile predictions. Those quantiles are calibrated as described in \cite{Romano2019_cqr}, to guarantee valid coverage, but they don't have guarantees on individual samples. Nevertheless, the difference between quantiles, or confidence interval (CI), can be seen as a proxy for the sample-wise uncertainty given by the model. This can be seen qualitatively in Figure \ref{fig:true_vs_pred}, where samples with high normalized CI are the ones that have higher error with respect to the true period. Here we test this approximation from a different perspective, looking at $T_{\mathrm{eff}}-P_{\mathrm{rot}}$ relationships and period distributions. Figure \ref{fig:teff_prot_ci80} shows  $T_{\mathrm{eff}}$ vs. $P_{\mathrm{rot}}$ at four different filtering schemes. Each filtering is done by posing a limit on the maximum $\text{CI}_{80}/P_{\mathrm{rot}}$ and removing samples higher than this limit. Each panel is also colored by $\text{CI}_{80}/P_{\mathrm{rot}}$. We can see that using the entire main-sequence population (left panel), there is a dark bulk of samples with high $\text{CI}_{80}/P_{\mathrm{rot}}$ and low periods. This bulk shows different $T_{\mathrm{eff}}-P_{\mathrm{rot}}$ behavior compared to the general population, as they are fast rotators with mild temperature. One of the possibilities for such fast rotators is that they are synchronized binaries (e.g., \cite{Simonian_2019} and \cite{Kamai2025_too_fast}). However, in this regime, there are actually two populations - large population with high  $\text{CI}_{80}/P_{\mathrm{rot}}$, and small population with low  $\text{CI}_{80}/P_{\mathrm{rot}}$, suggesting, that the high-CI population might have wrong periods. CI-based Filtering gradually removes them, without affecting the general population, and at a cutoff of $\text{CI}_{80}/P_{\mathrm{rot}} < 0.4$, the entire dark bulk is removed, keeping only high-confidence samples in the synchronized binaries regime. \\
Another test for confidence intervals as uncertainty is shown in Figure \ref{fig:prot_dist_compare}, where we compare the resulting $P_{\mathrm{rot}}$ distribution with the ones from \cite{McQuillan2014_acf_catalog} and \cite{Santos2021_catalog}. It can be seen that using the full main-sequence dataset, the distribution of our model (blue) is not consistent with \cite{McQuillan2014_acf_catalog} and \cite{Santos2021_catalog}, and shows a peak at short periods. Using the same $\text{CI}_{80}/P_{\mathrm{rot}} < 0.4$ cutoff, the three distributions match, and the median $P_{\mathrm{rot}}$ of our filtered distribution ($17.94$ days) is between the medians of \cite{McQuillan2014_acf_catalog} and \cite{Santos2021_catalog}.\\
The two tests suggest that the confidence intervals predicted by the model can be used as an approximation to the model's uncertainty, at least for the filtering procedure. Moreover, it seems like $0.4$ is a reasonable upper limit on the $80\%$ normalized CI, keeping $119{,}428$ ($\sim 80\%$) samples that agree with previous catalogs. Of these, $110{,}145$ have a $T_{\rm eff}$ from \cite{Berger2020_kepler_prop} and enter the $T_{\rm eff}$--$P_{\rm rot}$ diagrams (Figs.~\ref{fig:teff_prot_ci80} and \ref{fig:teff_prot_gyro}), which shows the expected $T_{\mathrm{eff}}-P_{\mathrm{rot}}$ behavior. Nevertheless, the $80\%$ CI is not the only uncertainty proxy; the exact filtering value might depend on the scientific task. 

\subsection{Comparison with previous catalogs}\label{sec:compare_cat}
\begin{figure*}
    \centering
    \includegraphics[width=0.5\linewidth]{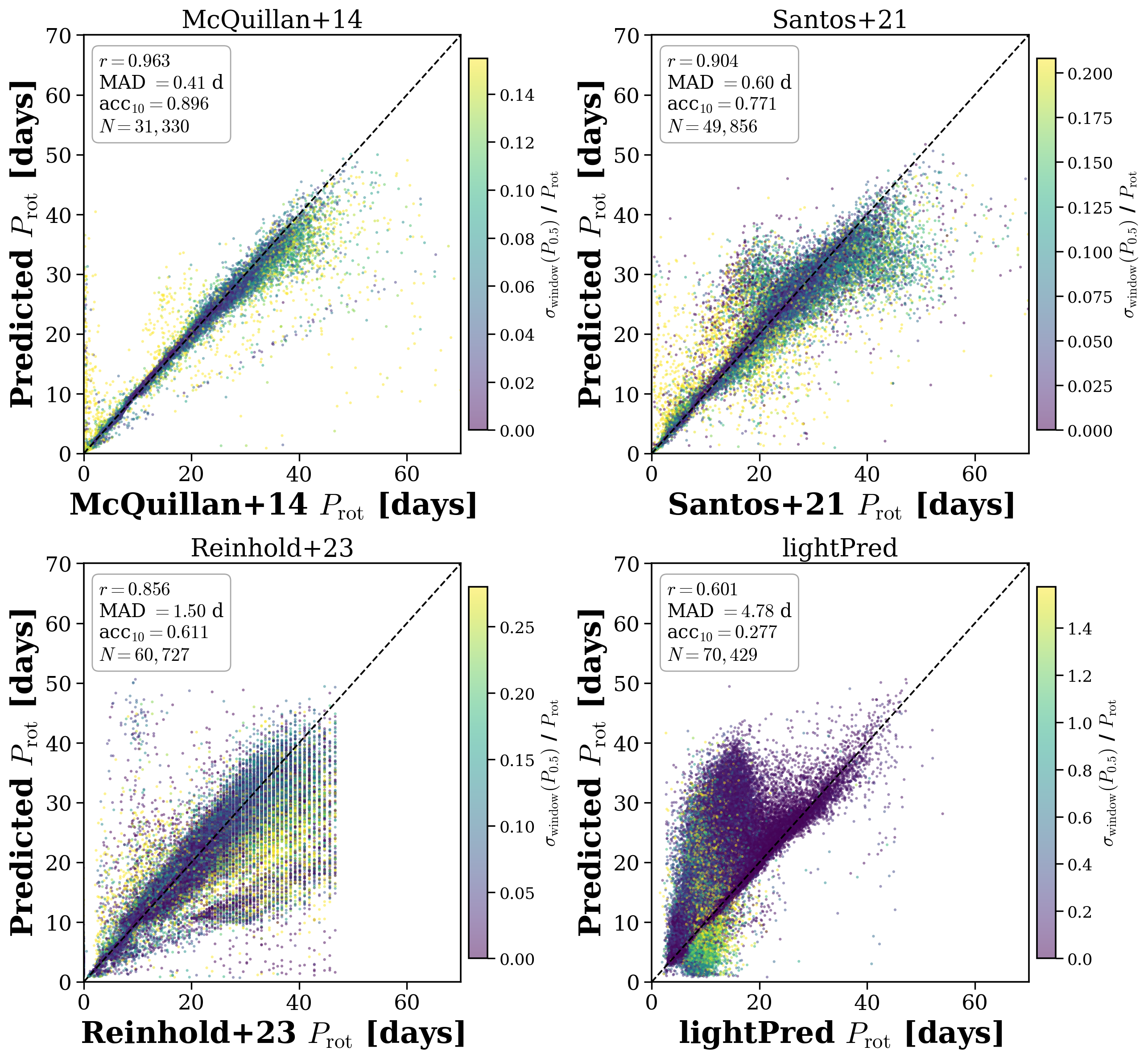}
    \caption{comparison between our periods and previous catalogs Color represents the normalized standard deviation of window predictions.}\label{fig:compare_catalogs}
\end{figure*}

As our training labels are only consensus periods, it is interesting to compare the full previous catalogs with our predictions. Figure \ref{fig:compare_catalogs} shows such a comparison with \cite{McQuillan2014_acf_catalog}, \cite{Santos2021_catalog}, \cite{Reinhold2023_gps_catalog}, and \cite{Kamai2025_lightpred}. The figure reveals several interesting observations. First, the agreement decreases with the size of the catalog. This is expected, as we already showed that larger catalogs include more samples with low coherence. Second, we see a qualitative difference between \cite{Kamai2025_lightpred} and the other catalogs. In all other catalogs, there is a strong correlation between agreement and $\sigma_{\mathrm{win}}$, which is expected. In the comparison with \cite{Kamai2025_lightpred}, we don't see such a strong correlation, and the total agreement is much lower than with other catalogs. As \cite{Kamai2025_lightpred} used a model that was trained on simulations, this highlights the gap between simulations and real data. It is important to emphasize that this gap does not affect our training labels, as we require consensus labels. This means that the only labels from \cite{Kamai2025_lightpred} that are used as labels are the ones that agree with at least one additional catalog.  

\subsection{Stellar Age}\label{sec:age}
\begin{figure*}
    \centering
    \begin{minipage}{0.49\linewidth}
    \includegraphics[width=\linewidth]{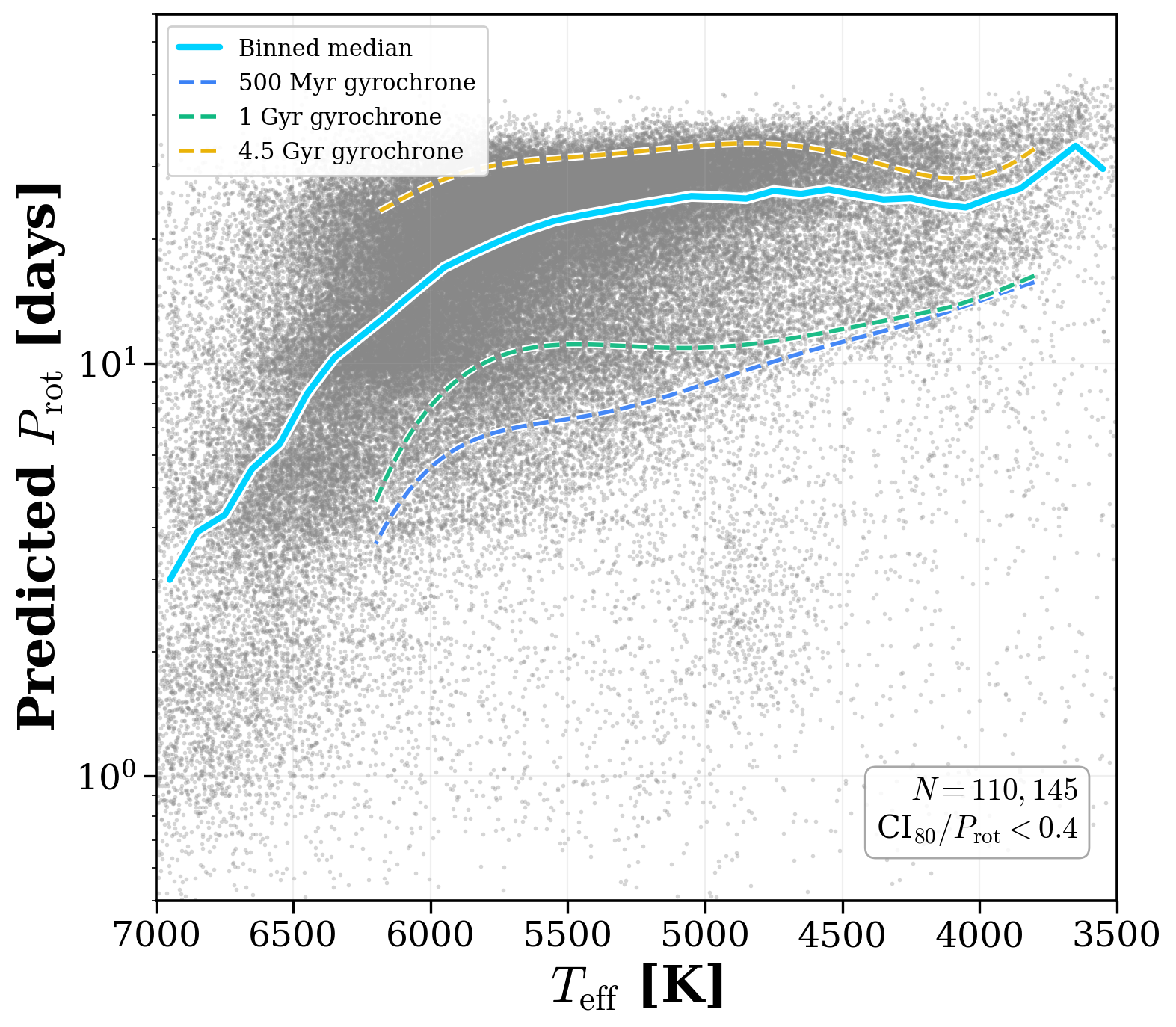}
    \end{minipage}
    \begin{minipage}{0.49\linewidth}
    \includegraphics[width=\linewidth]{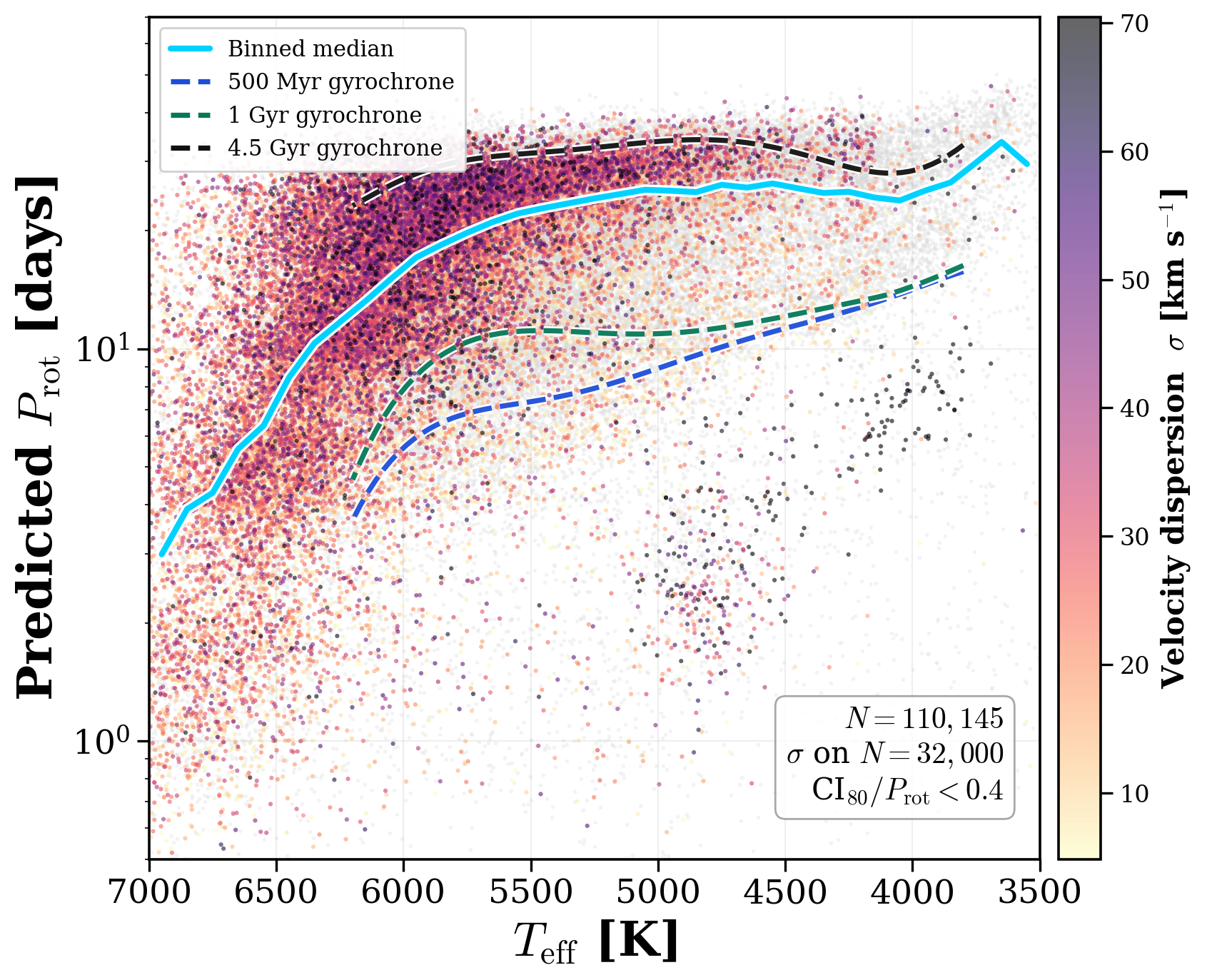}
    \end{minipage}
    \caption{$T_{\mathrm{eff}}$ vs. $P_{\mathrm{rot}}$ of main sequence dataset filtered with $\text{CI}_{80}/P_{\mathrm{rot}}<0.4$. The dashed lines represent gyrochronology curves at different ages, using the model from \cite{bouma_empirical_2023}. The difference between the panels is the coloring - the left panel uses no colors, and the right panel uses velocity dispersion, $\sigma_\star$, as color (samples with no $\sigma_\star$ are excluded from the right panel) }\label{fig:teff_prot_gyro}
\end{figure*}

In the left panel of Figure \ref{fig:teff_prot_gyro} we show $T_{\mathrm{eff}}$ vs. $P_{\mathrm{rot}}$ for the main sequence predictions with the suggested cutoff of $\text{CI}_{80}/P_{\mathrm{rot}}<0.4$. We overlay three empirical gyrochronology curves from \cite{bouma_empirical_2023}. It can be seen that the $4.5$ Gyr curve marks the upper envelope of the scatter, as expected from the Kepler population (see, for example, \cite{bouma_ages_2024}). At the lower end, we see a non-negligible population below $500$ Myr. As some of them are young populations, the others might be synchronized binaries. The right panel shows the same figure, now colored by the per-star kinematic-age proxy
$\sigma_\star \equiv \sqrt{(U^2+V^2+W^2)/3}$, where $U,V,W$ are the Galactic
space-velocity components relative to the local standard of rest, from Gaia DR3
astrometry and radial velocities. Stellar populations are dynamically heated
over time, so their velocity dispersion grows with age
\citep{Spitzer1951_kinematic, Nordstrom2004_kinematic}; $\sigma_\star$ is the
single-star analogue, whose ensemble root-mean-square recovers the per-axis
dispersion of a population, and larger values therefore indicate an on average
older population.

\begin{figure*}
    \centering
    \includegraphics[width=\linewidth]{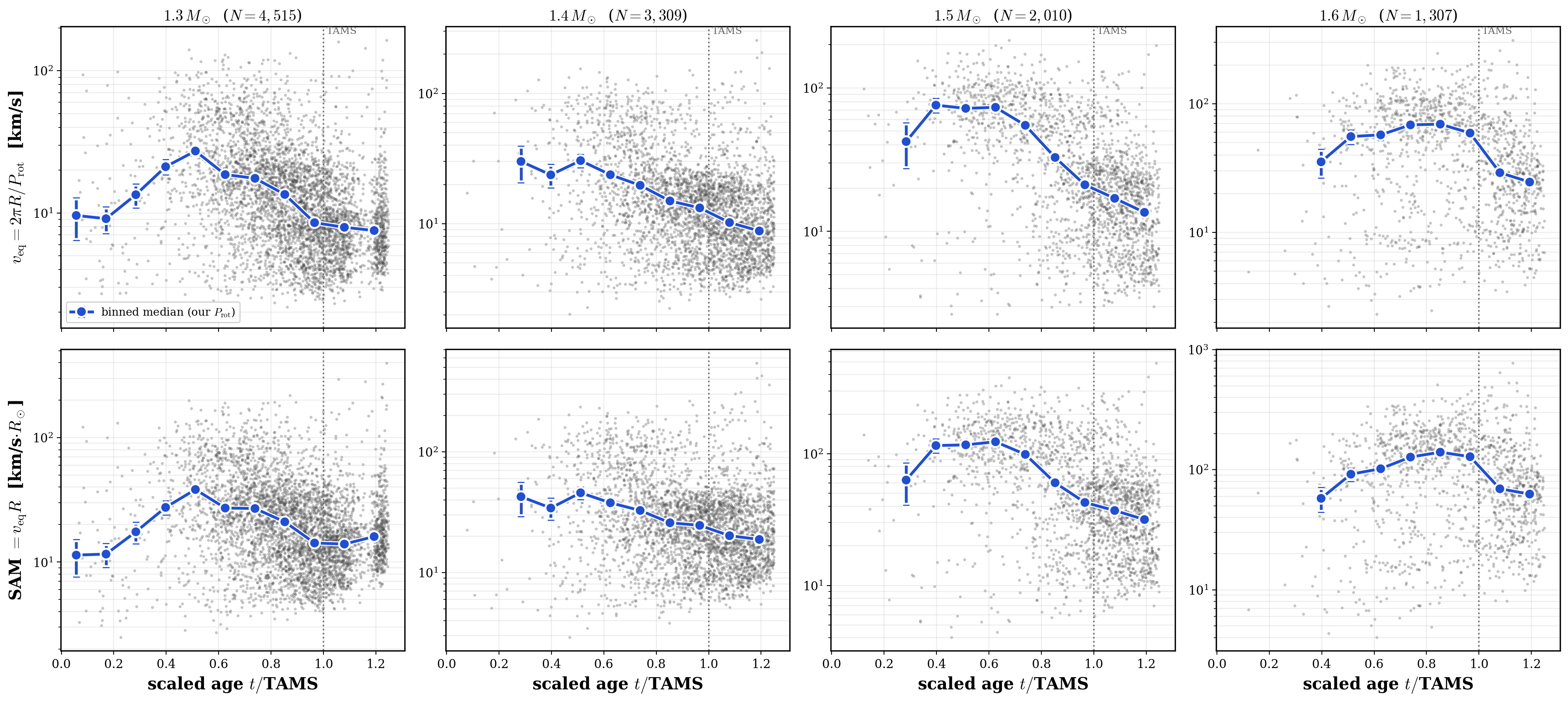}
    \caption{scaled age vs. equatorial velocity (upper row) and SAM (lower row). Each panel shows a different mass bin. In each panel, the blue line is the median over scaled age bins. }\label{fig:veq_vs_scaled_age}
\end{figure*}

Another interesting investigation is the analysis of angular momentum as a function of time. We expect a qualitative difference between high-mass stars above the Kraft break and low-mass stars. This difference was shown in \cite{Sussholz2026_age_vbroad} for projected velocity, but since we have rotation periods, we can combine it with the radius and implement the same test for equatorial velocity $v_{\mathrm{eq}}=\frac{2\pi R}{P_{\mathrm{rot}}}$. This cleans the $sin(i)$ projection factor and serves as a physical test for our predictions.\\
To carry out this test, we cross-match our main-sequence catalogue against Gaia DR3, taking radii, masses, and ages from the FLAME module of \texttt{gaiadr3.astrophysical\_parameters}, and effective temperatures
(together with a radius-quality proxy) from GSP-Phot. KIDs are bridged to
Gaia DR3 \texttt{source\_id} through the cross-match of
\citet{Godoy-Rivera2025}. For each star we convert the
photometric period to an equatorial velocity, $v_{\rm eq} = \frac{2\pi R}{P_{\rm rot}}
           = 50.6\,\frac{R/R_\odot}{P_{\rm rot}/{\rm d}}\ {\rm km\,s^{-1}}$,
using the FLAME radius, together with a specific-angular-momentum (SAM) tracer
$j \equiv v_{\rm eq}\,R$.
We adopt the age normalization of \citet{Sussholz2026_age_vbroad}: the evolutionary
coordinate is the scaled age $t/\mathrm{TAMS}(M)$, with the terminal-age
main sequence defined as the age at which the central hydrogen fraction
first falls below $X_{\rm cen} < 2\times10^{-4}$. We tabulate
$\mathrm{TAMS}(M)$ from the YREC grids from
\texttt{kiauhoku} \citep{Claytor2020_kiauhoku, claytor2025}, interpolated with a monotonic cubic over $0.68 \le M/M_\odot \le 2.0$. Dividing
by $\mathrm{TAMS}(M)$ removes the first-order mass dependence of the
main-sequence lifetime, placing stars of different mass on a common
evolutionary axis.
We then apply the selection of \citet[][Sec.~2]{Sussholz2026_age_vbroad}:
$4000 < T_{\rm eff} < 8000$~K, parallax S/N $>10$, and FLAME/GSP-Phot
radius and mass S/N $>5$.
Figure~\ref{fig:veq_vs_scaled_age} uses the four mass columns straddling the Kraft
break, $M = 1.3, 1.4, 1.5, 1.6\,M_\odot$, and plots scaled age vs. $v_{\mathrm{eq}}$ (upper row) and SAM (lower row). We see the same picture found in \cite{Sussholz2026_age_vbroad}; across the four bins the characteristic $v_{\rm eq}$ and SAM rise steeply with mass,
from $\sim\!10$--$20\ {\rm km\,s^{-1}}$ at $1.3\,M_\odot$ to $\sim\!10^2\ {\rm km\,s^{-1}}$
at $1.6\,M_\odot$, as expected from the progressive weakening of magnetic braking above the Kraft break. In addition, at $1.6\,M_\odot$, both $v_{\rm eq}$ and SAM are relatively flat up until $\tau/\text{TAMS}\sim 1$.

\subsection{Metallicity dependency}
\label{sec:feh}

\begin{figure*}
\centering
\includegraphics[width=0.75\textwidth]{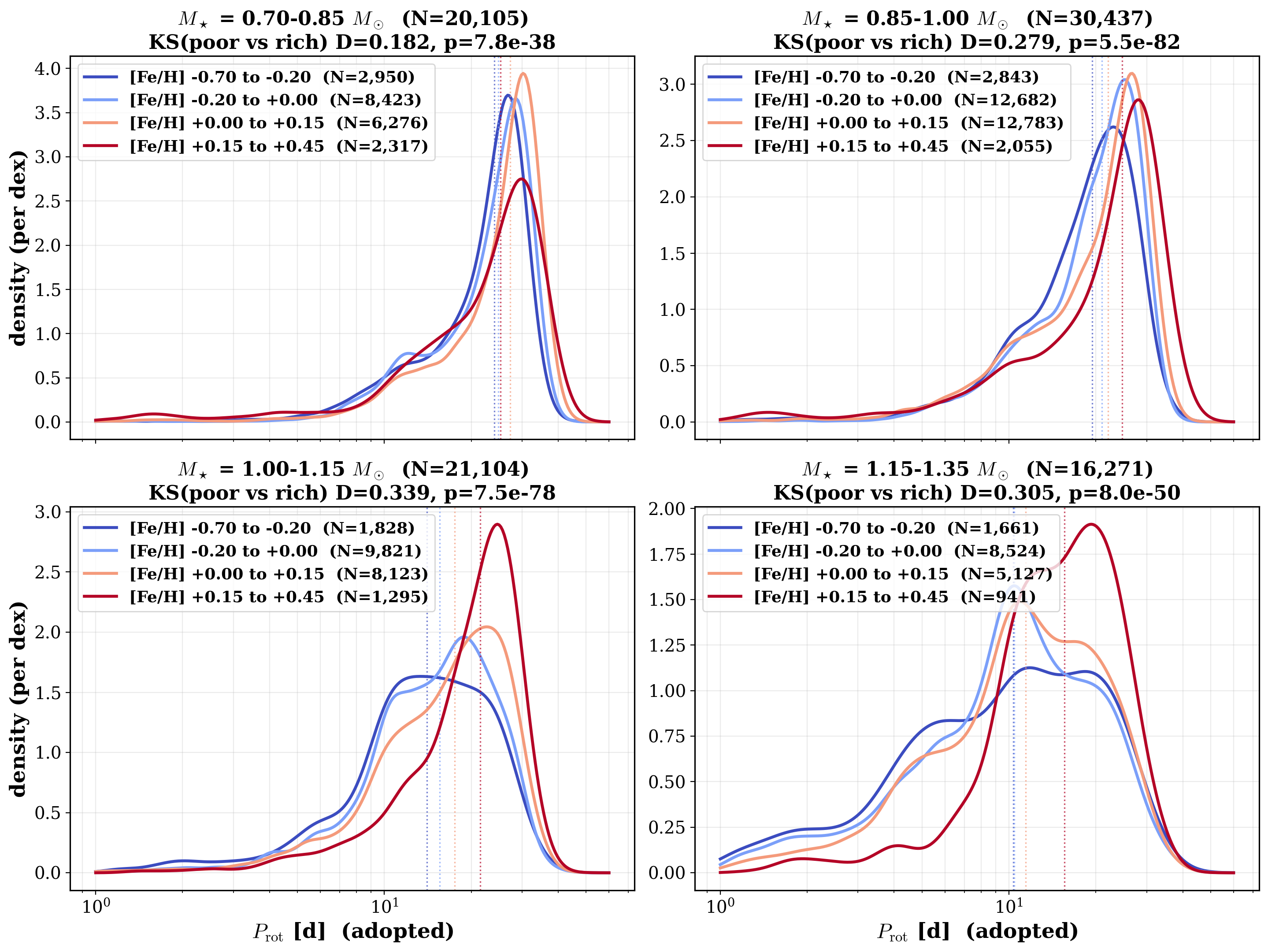}
\caption{Adopted-$P_\mathrm{rot}$ distributions ($\mathrm{CI}_{80}/P_\mathrm{rot}<0.4$)
in four $\mathrm{[Fe/H]}$ intervals, split into four stellar-mass bins. Vertical
dotted lines mark the per-interval medians; legends give sample sizes and median
periods, and each panel lists the metal-poor vs.\ metal-rich KS statistic.
Masses and metallicities are from \citet{Berger2020_kepler_prop}. Above $0.85\,M_\odot$ the
median period increases with metallicity at fixed mass, as expected from
metallicity-dependent magnetic braking.}
\label{fig:feh_massbins}
\end{figure*}

We now recover the metallicity dependence of rotation at
fixed stellar mass. Metallicity is expected to affect rotation through a metallicity-dependent magnetic braking: at
fixed mass, higher $\mathrm{[Fe/H]}$ deepens the convective envelope and
lengthens the convective-turnover time, lowering the Rossby number and
strengthening the wind torque, so metal-rich stars spin down to longer periods
\citep{vansaders_2013, amard_matt_2020}. This effect was observed by \cite{See2024_feh_prot} on a smaller sample, and here we test it on our sample. 
We cross-match the filtered catalogue
($\mathrm{CI}_{80}/P_\mathrm{rot}<0.4$) against the masses and metallicities of
\citet{Berger2020_kepler_prop} and split it into four mass bins over $0.70$--$1.35\,M_\odot$;
within each bin we compare the adopted-$P_\mathrm{rot}$ distributions in four
$\mathrm{[Fe/H]}$ intervals (Figure~\ref{fig:feh_massbins}).
Above $0.85\,M_\odot$ the median rotation period increases monotonically with
metallicity. At $1.0$--$1.15\,M_\odot$ the median rises from
$14.0$\,d for $\mathrm{[Fe/H]}\in[-0.7,-0.2]$ to $21.5$\,d for
$\mathrm{[Fe/H]}\in[+0.15,+0.45]$, and the metal-poor and metal-rich
distributions differ at high significance in every bin (two-sample KS
$D=0.18$--$0.34$, $p<10^{-37}$). The offset grows with mass, from $\sim1$\,d at
$0.70$--$0.85\,M_\odot$ to $5$--$7$\,d above solar mass. We note that this trend cannot be explained by the age--metallicity relation of
the Galactic disk. Chemical enrichment means that metal-rich stars are on
average younger, and gyrochronology requires younger stars to rotate faster, so
an age effect alone predicts that median $P_{\rm rot}$ should decrease
with $\mathrm{[Fe/H]}$ at fixed mass, the opposite of what we observe. We
verify that the age ordering runs in the expected direction using velocity
dispersion, $\sigma_\star$, as an independent age proxy (see Section~\ref{sec:age}). Figure~\ref{fig:feh_prot_vdisp} (Appendix~\ref{sec:appendix})
compares the two effects in the same mass bins: median $P_{\rm rot}$ rises with
$\mathrm{[Fe/H]}$ (left panel) while $\sigma_\star$ falls (right panel), confirming
that the metal-rich stars in our sample are indeed the kinematically colder,
younger ones. 

\subsection{Binaries}
\begin{figure*}
    \centering
    \begin{minipage}{0.49\linewidth}
    \includegraphics[width=\linewidth]{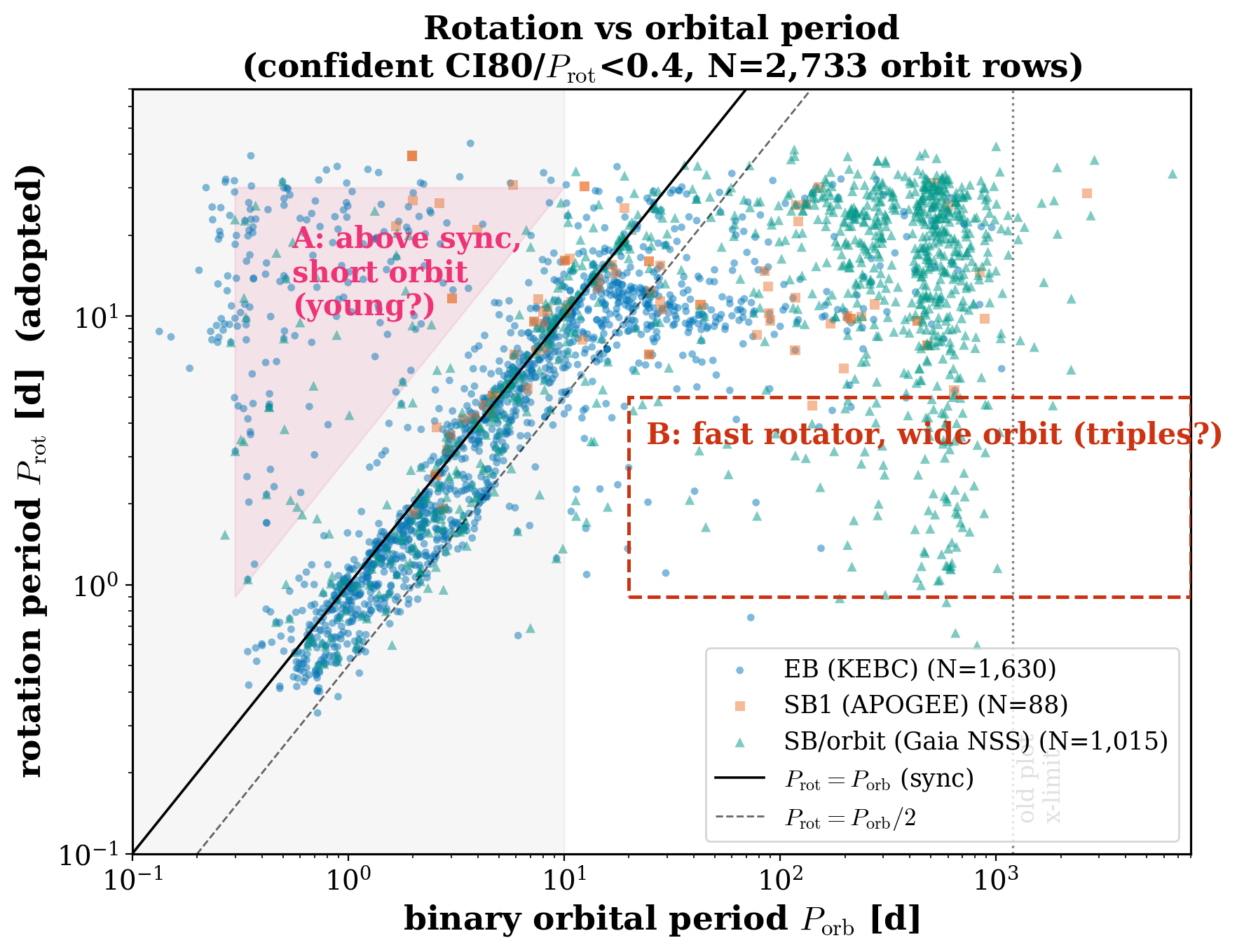}
    \end{minipage}
    \begin{minipage}{0.49\linewidth}
    \includegraphics[width=\linewidth]{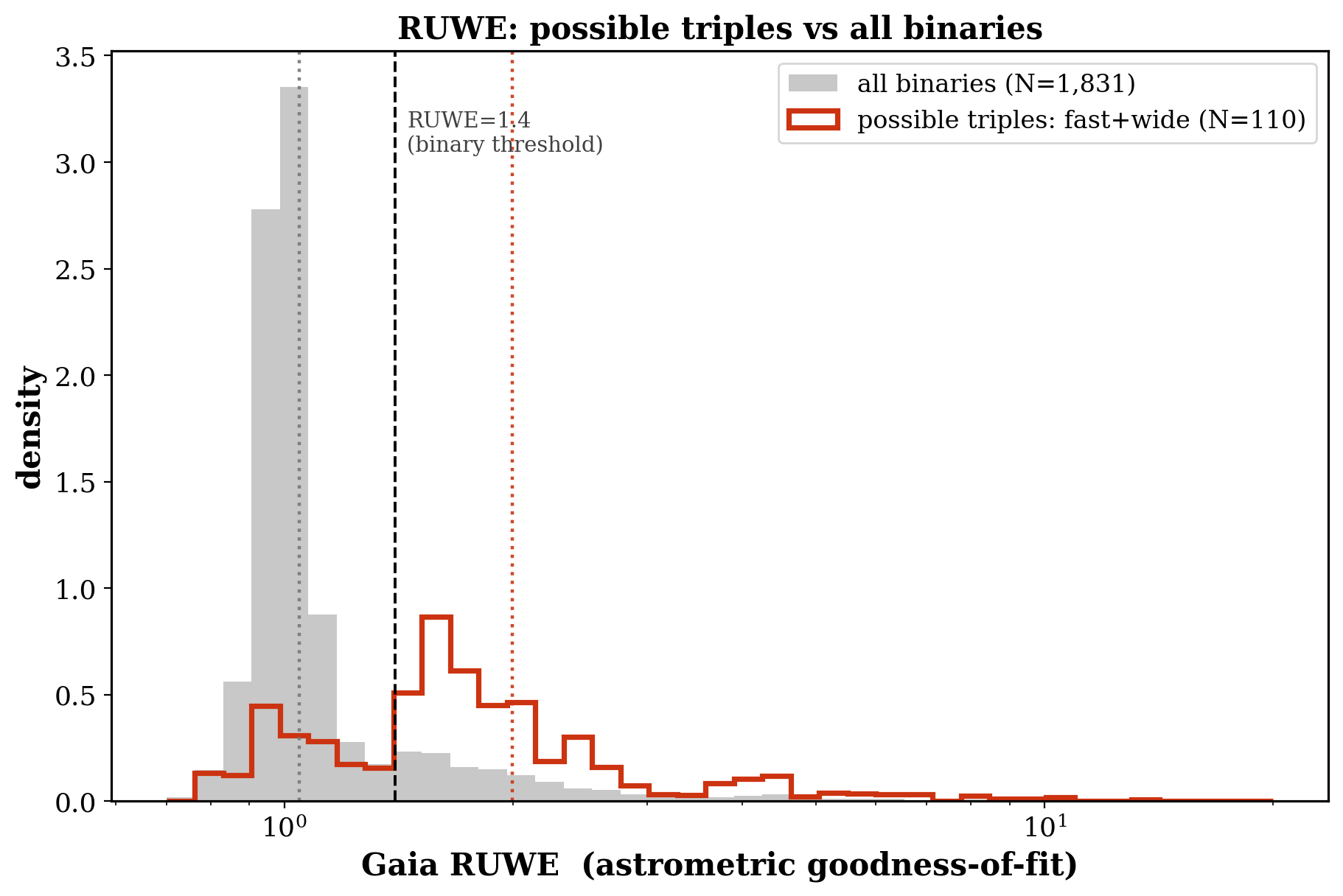}
    \end{minipage}
    \caption{Binaries in the \emph{Kepler} field. Left - orbital vs. stellar period. Different colors and shapes represent different types of binaries. Regime A and regime B are detailed in the text. Right - Renormalized Unit Weighted Error (RUWE) of the entire binary population (gray) and samples in regime B from the left panel (red).}\label{fig:binaries}
\end{figure*}

We finish with an analysis of known binaries. Figure~\ref{fig:binaries} shows on the left panel the orbital period vs. the predicted rotation period, for confident samples ($CI_{80}/P_{\rm{rot}}<0.4$). Tidal synchronization is a well-known phenomenon, already observed multiple times in Kepler \citep{Lurie_2017, Simonian_2019, Kamai2025_lightpred}. The gap around one year is also expected from the Gaia selection bias. However, we find several interesting regimes in the figure. The first is a regime where the orbital period is fast (below $10$ days) but the stellar period is not synchronized (regime A in the left panel of Figure~\ref{fig:binaries}. This might be a result of samples that are in the process of synchronization. If this is the case, we expect them to be younger and more eccentric compared to the synchronized population. We test the three-dimensional velocity dispersion of both the synchronized and
the above-sync (regime A) populations. Unlike the per-star proxy $\sigma_\star$
of Section~\ref{sec:age}, this is an ensemble quantity, measured about each
group's own median and summed over the three axes without the $1/3$ normalization, so the two are not numerically comparable. For every velocity component $x\in\{U,V,W\}$ we take the
median absolute deviation about the median, scaled by $1.4826$ to a Gaussian-equivalent width, and combine them in quadrature,
$\sigma_{\rm 3D}=1.4826\,\sqrt{\mathrm{MAD}_U^2+\mathrm{MAD}_V^2+\mathrm{MAD}_W^2}$.
a lower $\sigma_{\rm 3D}$ indicates a kinematically colder, on average younger,
population. Uncertainties are the $2.5$--$97.5$ percentile range from $2000$
bootstrap resamplings. We find that the above-sync population has $\sigma=36.8$ with an uncertainty interval of $[29.7, 48.3]$. The synchronized population has $\sigma=45.5$ with interval $[39.5, 51.1]$. The above-sync population shows lower $\sigma$, but the two values sit within the interval of each other, so we cannot conclude that regime A is younger. Looking at eccentricities from \cite{gaia:nss} and \cite{IJspeert2024_eb_ecc} we find similar picture. The median eccentricity of the above-sync population is $0.05$, the median eccentricity of the synchronized population is $0.03$, and both distributions match.\\
The next interesting regime is the opposite regime---orbital period is long but stellar period is short (regime B in the left panel of Figure~\ref{fig:binaries}). We suggest that this regime contains triple systems, and what we identify as a wide binary is in many cases the triple companion. We indeed see that the astrometric error, Renormalized Unit Weight Error (RUWE), is statistically higher in regime B, compared to the general binary population. We also find evidence for that in \cite{Borkovits2025_triples}, which analyzed hierarchical triple systems in \emph{Kepler}. There is only one system from their catalog in regime B---\emph{KID 6525196}. \cite{Borkovits2025_triples} characterized the triple period as $418$ days, exactly the 'binary' period in our data. They also characterized the inner binary period as $3.42$ days, while our prediction for the stellar period is $1.61$ days, a factor of $\sim 0.5$. This is consistent with the fast-rotating star being a close binary (with a factor of 2 difference between our prediction and \cite{Borkovits2025_triples}), and the long period being the triple companion, as we hypothesized. We conclude that samples in regime B are probably hierarchical triple systems, which a dedicated follow-up can test. \\
Beyond the two above-mentioned regimes, we also see a horizontal line between orbital periods of $\sim 30$ and $\sim 100$ days. We do not fully understand the nature of this line and cannot conclude if it's a physical or systematic phenomenon.

%% file: sections/catalog.tex
\section{Period Catalog}\label{sec:catalog}
We release the catalog in two forms. The primary product is a \emph{merged} per-star catalog (Table~\ref{tab:catalog}): one row per star, giving the adopted rotation period \texttt{adopted\_Prot}, its five calibrated quantiles, the two uncertainty metrics ($\sigma_{\rm win}$ and $\mathrm{CI}_{80}/P$), and the bimodality diagnostics of
Section~\ref{sec:bimodal}. The adopted period depends on the split type: for unimodal stars it is the median over rolling windows; for distinct bimodals it is the long-cluster
aggregate (\texttt{period\_source}\,=\,\texttt{long\_cluster}) identified as the true
rotation by the $v\sin i$ test, with the low mode retained as \texttt{mode\_low} and
the adopted long mode as \texttt{mode\_high}; for $2{:}1$ harmonics the window median is
kept and the star is flagged \texttt{mode\_ambiguous}. Stars with fewer than four windows,
for which bimodality cannot be assessed, are labelled \texttt{too\_few\_windows} and keep
the window median. We record both candidate modes
(\texttt{mode\_low}, \texttt{mode\_high}) and the adopted choice (\texttt{period\_source})
so users can revert or re-weight the assignment for their own tasks.

We additionally release the underlying \emph{window-level} catalog: one row per
$(\mathrm{KID},\ \text{window})$, giving the window start and the five calibrated period quantiles for that window. This is the raw product from which the per-star aggregates and the bimodality flags are derived, and it exposes the per-window period switching directly.
Table~\ref{tab:catalog_windows} shows an example for KID~892376, whose per-window period alternates between its $\sim1.4$~d and $\sim14$~d modes; the merged catalog adopts the long
($14.1$~d) mode for this star even though the raw window median ($1.52$~d) falls in the short mode, illustrating why the adjudication is necessary.

Because the calibrated interval and $\sigma_{\rm win}$ track reliability without imposing a
single quality cut, we publish the full catalog rather than a filtered one: the
$\mathrm{CI}_{80}/P_{\rm rot}<0.4$ selection is a reasonable default for population studies,
but analyses of failure modes, aliasing, or the ambiguous majority need the unfiltered
predictions. Some periods, particularly high-$\mathrm{CI}$ and ambiguous-bimodal stars, are unreliable, which is exactly why the per-star reliability metrics and the full mode diagnostics are provided. Both tables will be available in full online upon publication.

 \begin{deluxetable*}{rcccccccccccl}
  \tabletypesize{\scriptsize}
  \tablecaption{Example rows from the merged per-star catalog, spanning the period range and all
  four \texttt{split\_type} values. \texttt{adopted\_Prot} and the calibrated quantiles
  $q_{10}$--$q_{90}$ are in days; $\sigma_{\rm win}\equiv\mathrm{std}(\log_{10}P_{\rm win})$ is the
  dispersion of the per-window periods (in dex, unitless and symmetric in period ratio), and
  $\mathrm{CI}_{80}/P$ the normalized calibrated $80\%$ width. For bimodal stars
 \texttt{mode\_low}/\texttt{mode\_high} give the two cluster periods and \texttt{period\_source}
  the adopted choice. The catalog would be available online upon publication.\label{tab:catalog}}
  \tablehead{
  \colhead{KID} & \colhead{adopted} & \colhead{$q_{10}$} & \colhead{$q_{25}$} &
  \colhead{$q_{50}$} & \colhead{$q_{75}$} & \colhead{$q_{90}$} & \colhead{$\sigma_{\rm win}$} &
  \colhead{$\mathrm{CI}_{80}/P$} & \colhead{$\mathrm{mode}_{\rm lo}$} &
  \colhead{$\mathrm{mode}_{\rm hi}$} & \colhead{split\_type} & \colhead{period\_source} \\
  \colhead{} & \colhead{(d)} & \colhead{(d)} & \colhead{(d)} & \colhead{(d)} & \colhead{(d)} &
  \colhead{(dex)} & \colhead{} & \colhead{(d)} & \colhead{(d)} &
  \colhead{} & \colhead{}
  }
  \startdata
  12352712 &  0.33 &  0.26 &  0.29 &  0.33 &  0.36 &  0.38 & 0.033 & 0.374 & \nodata & \nodata & unimodal         & unimodal        \\
   4577860 &  5.00 &  4.71 &  4.85 &  5.00 &  5.12 &  5.27 & 0.008 & 0.113 & \nodata & \nodata & unimodal         & unimodal        \\
  11775432 & 12.00 & 11.22 & 11.61 & 12.00 & 12.35 & 12.77 & 0.018 & 0.128 & \nodata & \nodata & unimodal         & unimodal        \\
   8818439 & 16.95 & 15.90 & 16.45 & 16.95 & 17.40 & 17.92 & 0.009 & 0.119 & \nodata & \nodata & unimodal         & unimodal        \\
   6671587 & 20.00 & 18.75 & 19.36 & 20.00 & 20.51 & 21.04 & 0.013 & 0.115 & \nodata & \nodata & unimodal         & unimodal        \\
    892376 & 14.07 & 13.25 & 13.68 & 14.07 & 14.39 & 14.80 & 0.505 & 0.111 &  1.42   & 14.02   & distinct         & long\_cluster   \\
   3542117 & 23.50 & 21.69 & 22.64 & 23.50 & 24.28 & 25.23 & 0.205 & 0.151 &  8.71   & 23.58   & distinct         & long\_cluster   \\
  10618253 &  0.51 &  0.37 &  0.44 &  0.51 &  0.58 &  0.66 & 0.122 & 0.576 &  0.31   &  0.54   & distinct         & long\_cluster   \\
   7369484 &  7.01 &  5.77 &  6.40 &  7.01 &  7.67 &  8.57 & 0.196 & 0.400 &  3.78   &  8.21   & harmonic\_2x     & window\_median  \\
   4469225 & 11.58 &  9.66 & 10.63 & 11.58 & 12.59 & 13.96 & 0.009 & 0.372 & \nodata & \nodata & too\_few\_windows & unimodal        \\
  \enddata
  \end{deluxetable*}
\begin{deluxetable}{rcccccc}
\tabletypesize{\footnotesize}
\tablecaption{Window-level catalog excerpt for the bimodal star KID~892376 (11 windows).
Each row is one $450$-day window; $P_{10}$--$P_{90}$ are its calibrated period quantiles in
days. The per-window period sits in the $\sim1.1$--$1.8$~d mode for windows $0$--$6$ and jumps
to the $\sim13.6$--$14.3$~d mode for windows $7$--$10$ ($n_{\rm low}=7$, $n_{\rm high}=4$);
the merged catalog adopts the long mode ($14.07$~d, Table~\ref{tab:catalog}).\label{tab:catalog_windows}}
\tablehead{
\colhead{win.} & \colhead{start} & \colhead{$P_{10}$} & \colhead{$P_{25}$} &
\colhead{$P_{50}$} & \colhead{$P_{75}$} & \colhead{$P_{90}$} \\
\colhead{} & \colhead{(d)} & \colhead{(d)} & \colhead{(d)} & \colhead{(d)} & \colhead{(d)} & \colhead{(d)}
}
\startdata
0  &   0 &  1.07 &  1.17 &  1.28 &  1.37 &  1.47 \\
1  &  90 &  0.93 &  1.03 &  1.13 &  1.21 &  1.31 \\
2  & 180 &  1.08 &  1.18 &  1.29 &  1.38 &  1.48 \\
3  & 270 &  1.24 &  1.35 &  1.46 &  1.56 &  1.67 \\
4  & 360 &  1.29 &  1.40 &  1.52 &  1.61 &  1.72 \\
5  & 450 &  1.28 &  1.40 &  1.51 &  1.61 &  1.72 \\
6  & 540 &  1.56 &  1.69 &  1.82 &  1.93 &  2.05 \\
7  & 630 & 12.84 & 13.26 & 13.64 & 13.95 & 14.36 \\
8  & 720 & 13.25 & 13.68 & 14.07 & 14.38 & 14.80 \\
9  & 810 & 13.51 & 13.95 & 14.33 & 14.65 & 15.07 \\
10 & 836 & 13.24 & 13.68 & 14.07 & 14.39 & 14.81 \\
\enddata
\end{deluxetable}

%% file: sections/conclusions.tex
\section{Conclusions}
\label{sec:conclusions}

We introduced The Maunder, a data-driven model for stellar rotation
periods in the \textit{Kepler} field. The model is trained with a hybrid
objective: a self-supervised term applied to all $148{,}746$
main-sequence stars, and a supervised, conformalized quantile-regression
term applied to the $41{,}650$ stars carrying a cross-catalog consensus
period. Drawing supervised labels from agreement between at least two of
\citet{McQuillan2014_acf_catalog}, \citet{Santos2021_catalog},
\citet{Reinhold2023_gps_catalog}, and \citet{Kamai2025_lightpred}
suppresses the single-catalog systematics that limit models trained on
one catalog, while avoiding the simulation-to-reality gap incurred by
training on synthetic light curves. On the held-out consensus test set,
the model reaches an RMSE of $2.36$~days, and its predicted intervals are
well calibrated, with empirical coverage matching the nominal $80\%$ and
$50\%$ levels. Its median absolute error, $0.46$~days, is below the error
implied for an individual reference catalog by the mutual scatter of the
input catalogs, so the model is more precise than any single catalog it
learns from.

A central design goal was per-star reliability. Because classical
periodograms expose mechanistic diagnostics that machine-learning models
usually lack, we equip every prediction with two complementary
uncertainty estimates: the conformalized confidence interval, which is
statistically valid by construction, and $\sigma_{\rm win}$, the scatter
of predictions across sliding windows of the same light curve. Both track
prediction error, and both grow for ambiguous or noisy targets. Filtering
on the normalized interval ($\mathrm{CI}_{80}/P_{\rm rot} < 0.4$) yields
$119{,}428$ stars whose period distribution (median $17.94$~days) and
$T_{\rm eff}$--$P_{\rm rot}$ relation are consistent with
\citet{McQuillan2014_acf_catalog} and \citet{Santos2021_catalog}. Rather
than imposing a single quality cut, we release the full quantile
predictions and $\sigma_{\rm win}$ so that users can define selection
functions suited to their science.

Inferring the period on rolling windows reveals a property of the sample
that a single-pass period search cannot expose. For $21.5\%$ of the
main-sequence population, the per-window periods do not form a single
cluster but split into two well-separated modes; $79\%$ of these are
genuinely distinct periods and $21\%$ are $2{:}1$ harmonics. To determine
which mode is the true rotation, we build a hierarchical forward model of
APOGEE $v\sin i$ (following \citealt{MasudaWinn2020_hbm}); for distinct
bimodals, the long mode reproduces the observed $v\sin i$ distribution
while the short mode requires an unphysical excess of rapid rotators,
identifying the long mode as the rotation and the short mode as an alias.
We adopt the long mode for distinct bimodals in the released catalog,
retain both candidate modes for every star, and leave the $2{:}1$
harmonics, which $v\sin i$ cannot resolve, flagged as ambiguous. This
identifies a population of $\sim$26,000 stars whose rotation period a
conventional single-window analysis would systematically mis-assign to
the shorter alias.

At fixed stellar mass, the median rotation period increases with
metallicity for stars above $0.85\,M_\odot$, with an offset that grows
toward higher mass. The sign of the trend is opposite to that of the
metallicity--age relation produced by chemical enrichment, so it cannot be attributed to an age effect,
and instead reflects metallicity-dependent magnetic braking
\citep{amard_matt_2020, amard_2020}. This confirms on the full
\textit{Kepler} main sequence an effect previously reported on a smaller
sample \citep{See2024_feh_prot}.

Combining the predicted periods with stellar radii removes the $\sin i$
projection factor and yields equatorial velocities and specific angular
momenta directly, where previous work on this population was restricted
to projected velocities \citep{Sussholz2026_age_vbroad}. Across the Kraft break
both quantities rise steeply with mass over $1.3$--$1.6\,M_\odot$, as
expected from the progressive weakening of magnetic braking in stars with
shallow convective envelopes.

The catalog further recovers the empirical gyrochronology sequences, with
the $4.5$~Gyr isochrone tracing the upper envelope of the
$T_{\rm eff}$--$P_{\rm rot}$ distribution and a distinct population below
$500$~Myr, and reproduces the tidal-synchronization locus among known
binaries. Within that population we identify a regime of short rotation
periods at wide orbital separations whose elevated astrometric noise,
together with one system in common with \citet{Borkovits2025_triples}, suggests
these are hierarchical triples in which the rotating star is itself a
close binary.

Several limitations frame these results. First, the reported accuracy is measured on the consensus subset: the most
confidently periodic $\sim\!28\%$ of the sample. For the ambiguous majority,
which has no consensus label by construction, no direct error metric is
available, and the calibrated interval and $\sigma_{\rm win}$ are the only
reliability information we can provide. We note that conformal coverage is
guaranteed only for stars exchangeable with the calibration set, which consists
of consensus-labeled stars; on the unlabeled population the intervals remain our
best uncertainty estimate but carry no formal guarantee. Second, the reliability metrics do not capture the model's most
costly failures. The confident harmonic aliases of
Section~\ref{sec:performance} are predicted consistently across windows,
so $\sigma_{\rm win}$ is small and the calibrated interval narrow, and
$26$ of $28$ survive the recommended cut. For $16$ of the $28$, the
dominant ACF peak of the light curve is itself aliased: these are stars on
which the classical periodicity is ambiguous and the published catalogs
recover the rotation only through pipeline-specific vetting that our model
does not learn. A power-based harmonic diagnostic, comparing ACF and
periodogram power at $P$, $2P$ and $P/2$, is the natural remedy and is
left for future work. Third, consensus labels reduce but do not eliminate
systematics: biases shared across the input catalogs (like crowding, for
example) might still propagate into the labels. Future work will extend
the approach to other surveys, in particular the shorter, noisier
\textit{TESS} baselines, and to multimodal settings combining
\textit{Kepler} light curves with different surveys. The released periods
and their calibrated uncertainties support rotation-based age inference
and spin-orbit studies across the full \textit{Kepler} main sequence.

%% file: sections/appendix.tex
\section{Appendix}\label{sec:appendix}

\subsection{Contamination and eclipsing-binary tests}\label{sec:appendix_blend}
Eclipsing binaries were identified by KID match against the Kepler Eclipsing
Binary Catalog \citep{Kirk2016_keplerEB}, using the \texttt{morph} parameter
to separate detached ($\texttt{morph}\rightarrow0$) from contact
($\rightarrow1$) systems; candidate synchronized binaries
($P_{\rm rot}<7$~d) were removed. Rates are quoted with Wilson intervals and
compared by Fisher exact test.

\texttt{CROWDSAP} values were read per quarter from the headers of the public
Kepler long-cadence light curves for quarters $3$--$16$. Bimodal and unimodal
samples of $700$ stars each were matched by nearest neighbour in
($T_{\rm eff}$, $K_p$). For each star we recorded the median
\texttt{CROWDSAP} across quarters and the spread
($\max-\min$); classes were compared with the Mann--Whitney $U$ test.

\subsection{Models comparison}

\begin{table}[!ht]
\centering
\caption{Self-supervised objective comparison on the held-out main-sequence
consensus test set ($n=4{,}187$). $R^2$ is on $\log_{10}P_{\mathrm{rot}}$;
$\mathrm{acc}@20$ is the fraction with relative period error below $20\%$;
RMSE is in days. The duality loss
\citep{Kamai2025_desa} matches the coupled-view DualFormer architecture
(Sec.~\ref{sec:model}); VICReg \citep{Bardes2021_vicreg} is included as a
standard joint-embedding baseline.}
\label{tab:ssl_ablation}
\begin{tabular}{lccc}
\toprule
SSL objective & $R^2$\,(log) & $\mathrm{acc}@20$ & RMSE\,(d)  \\
\midrule
Duality loss & \textbf{0.973} &  \textbf{0.952} & \textbf{2.36}  \\
VICReg      & 0.958 & 0.947 & 2.79  \\
\bottomrule
\end{tabular}
\end{table}

\begin{table}[!ht]
\centering
\caption{Comparison between our model, a time-only, and a frequency-only model. The time-only and frequency-only models are identical to the time and frequency branches in the full model.}
\label{tab:ssl_ablation_time_freq}
\begin{tabular}{lccc}
\toprule
Model & $R^2$\,(log) & $\mathrm{acc}@20$ & RMSE\,(d) \\
\midrule
Full model & \textbf{0.973} & \textbf{0.952} & \textbf{2.36}  \\
Time-only model       & 0.971  & 0.946 & 2.46  \\
Frequency-only model       & 0.954  & 0.947 & 2.49  \\
\bottomrule
\end{tabular}
\end{table}

\subsection{Additional figures}

\begin{figure}[!ht]
    \centering
    \includegraphics[width=0.5\linewidth]{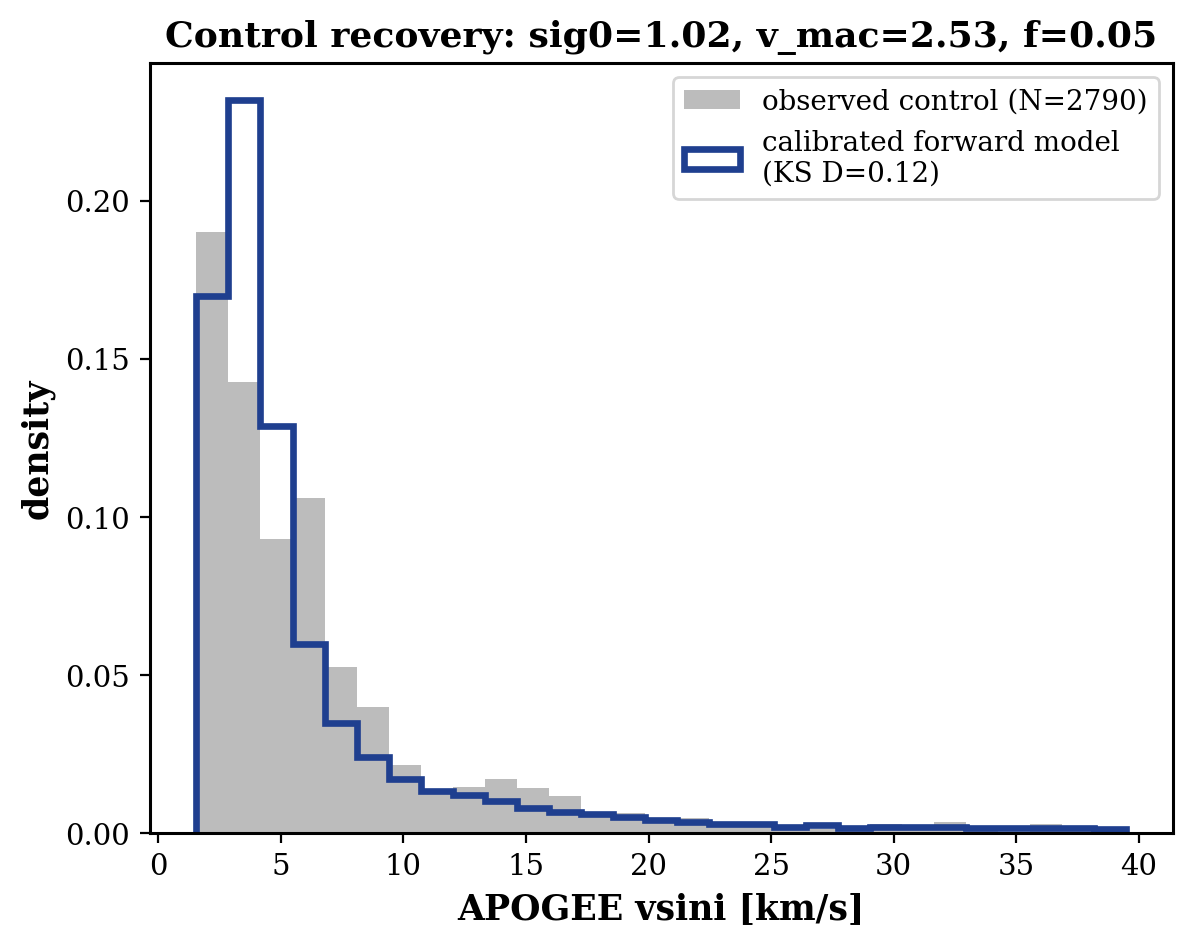}
    \caption{Calibration of the $v\sin i$ forward model on the unimodal control ($N=2790$). Under isotropic inclinations, the model with $\sigma_0 = 1.02$, $v_{\rm mac} = 2.53$ km/s (macroturbulence and measurement floor), and a fixed multiplicative error $f=0.05$, reproduces the observed APOGEE $v\sin i$ of stars with consistent (unimodal) periods (grey) to KS $D=0.12$. This frozen noise model is then applied unchanged to the bimodal mode comparison.}\label{fig:vsini_hbm_control}
\end{figure}

\begin{figure*}
    \centering
    \includegraphics[width=0.5\linewidth]{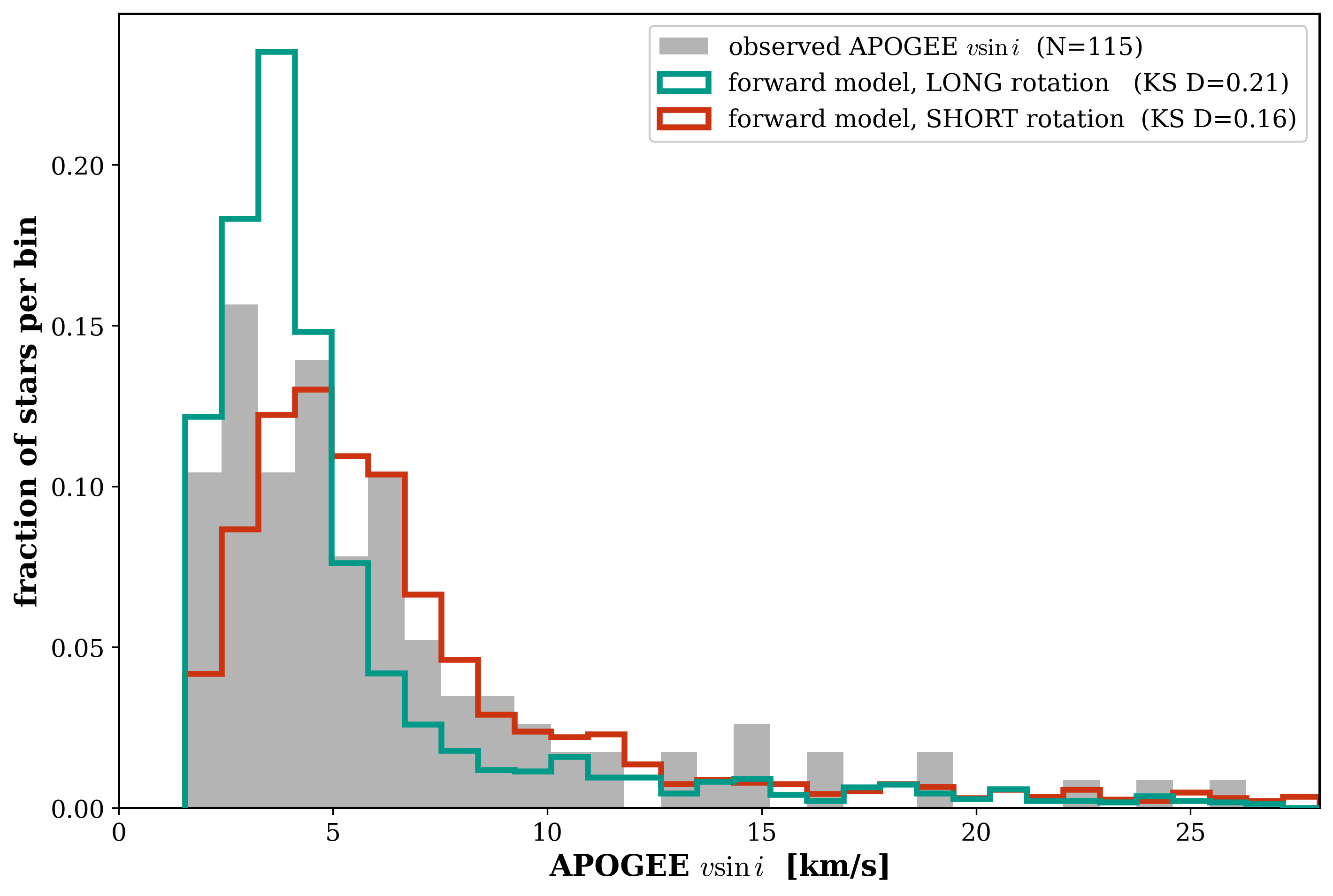}
    \caption{same as Figure~\ref{fig:vsini_hbm}, but for bimodals with 2:1 ratio}\label{fig:vsini_hbm_harmonic}
\end{figure*}

\begin{figure*}
    \centering
    \includegraphics[width=0.7\linewidth]{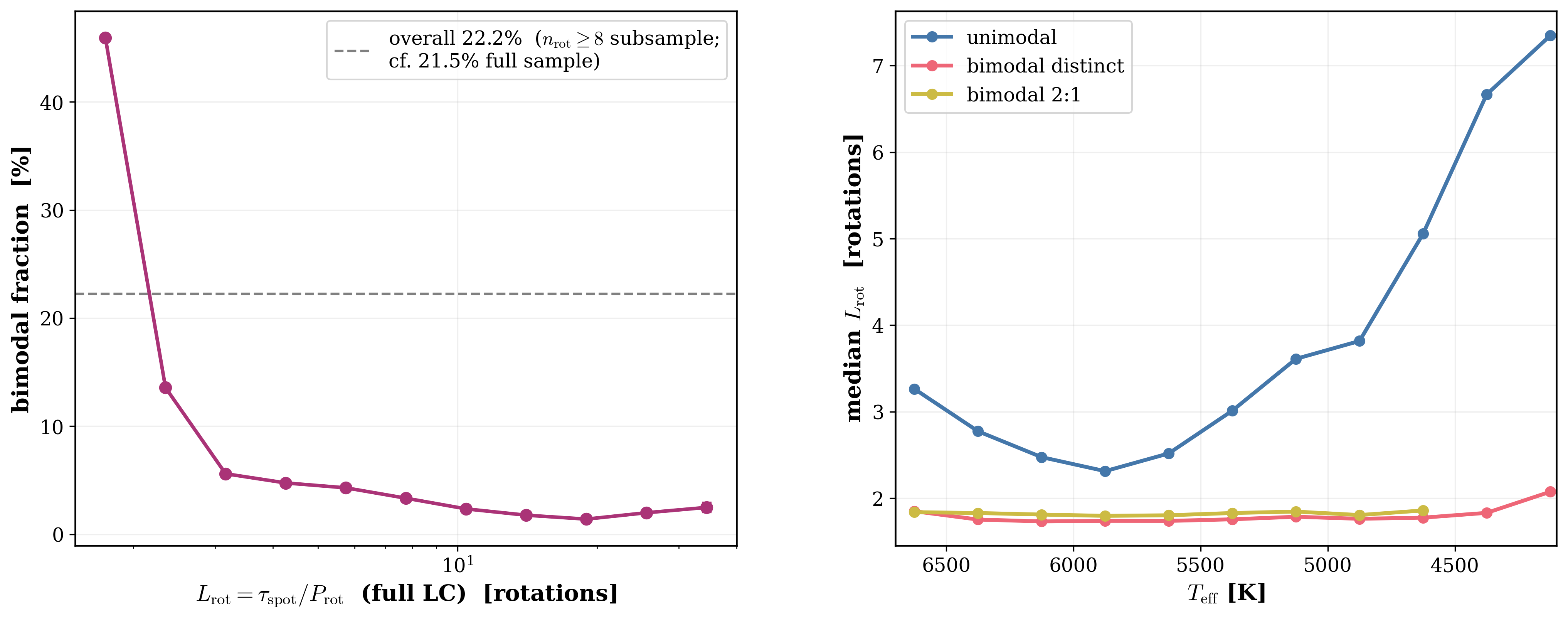}
    \caption{same as Figure ~\ref{fig:L_day}, but for $L_{\rm{rot}}$.}\label{fig:L_rot}
\end{figure*}

\begin{figure*}
    \centering
    \includegraphics[width=0.7\linewidth]{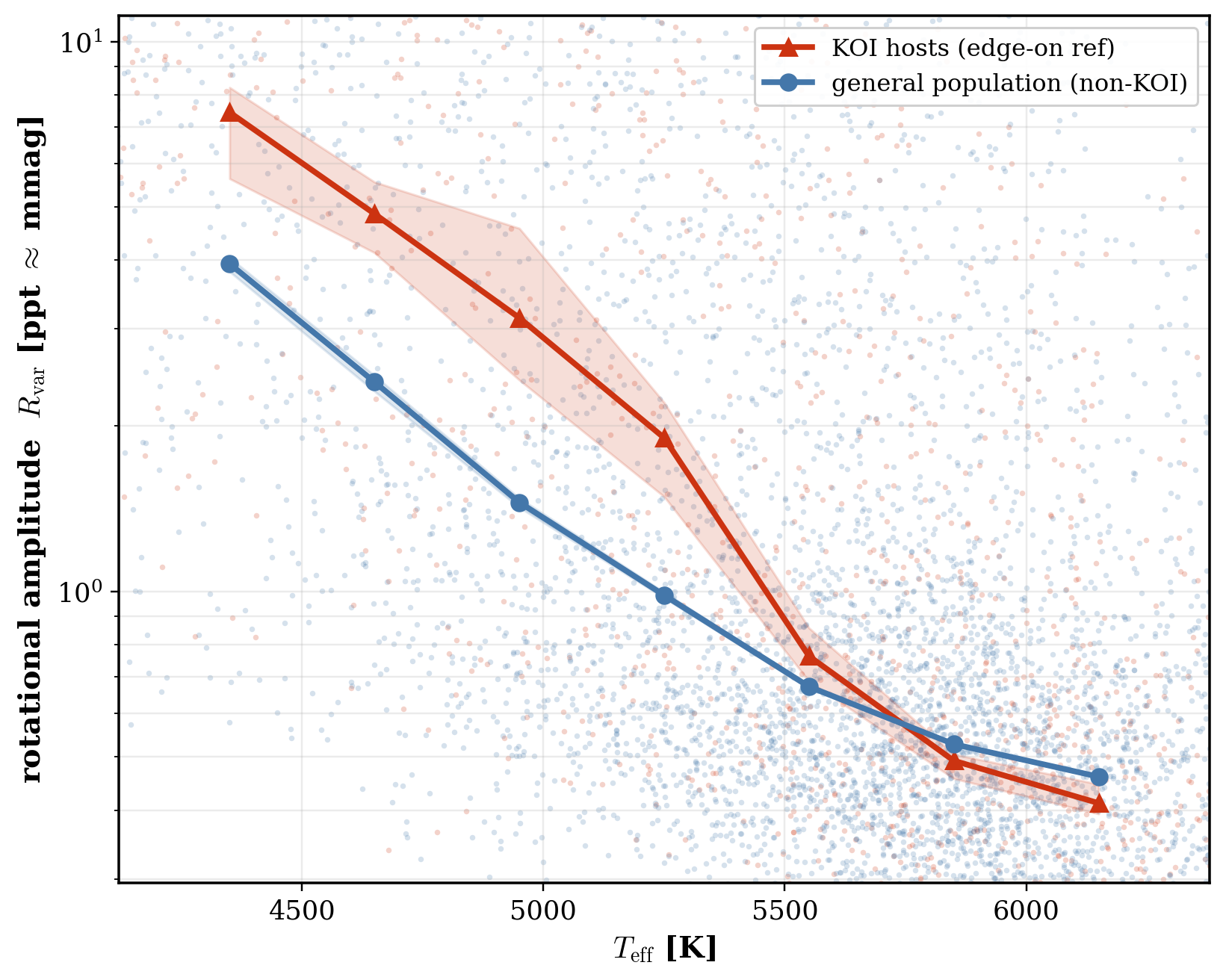}
    \caption{Photometric amplitude vs. $T_{\rm{eff}}$ for planet hosts (red) and non-planet hosts (blue). This is a reproduction of Figure 1 in \cite{Mazeh2015_inclination}. Error bars are estimated from bootstrapping - resample each bin with replacement and report one $\sigma$ interval of the medians. }\label{fig:inclination_mazeh_reproduce}
\end{figure*}

\begin{figure*}
    \centering
    \includegraphics[width=0.7\linewidth]{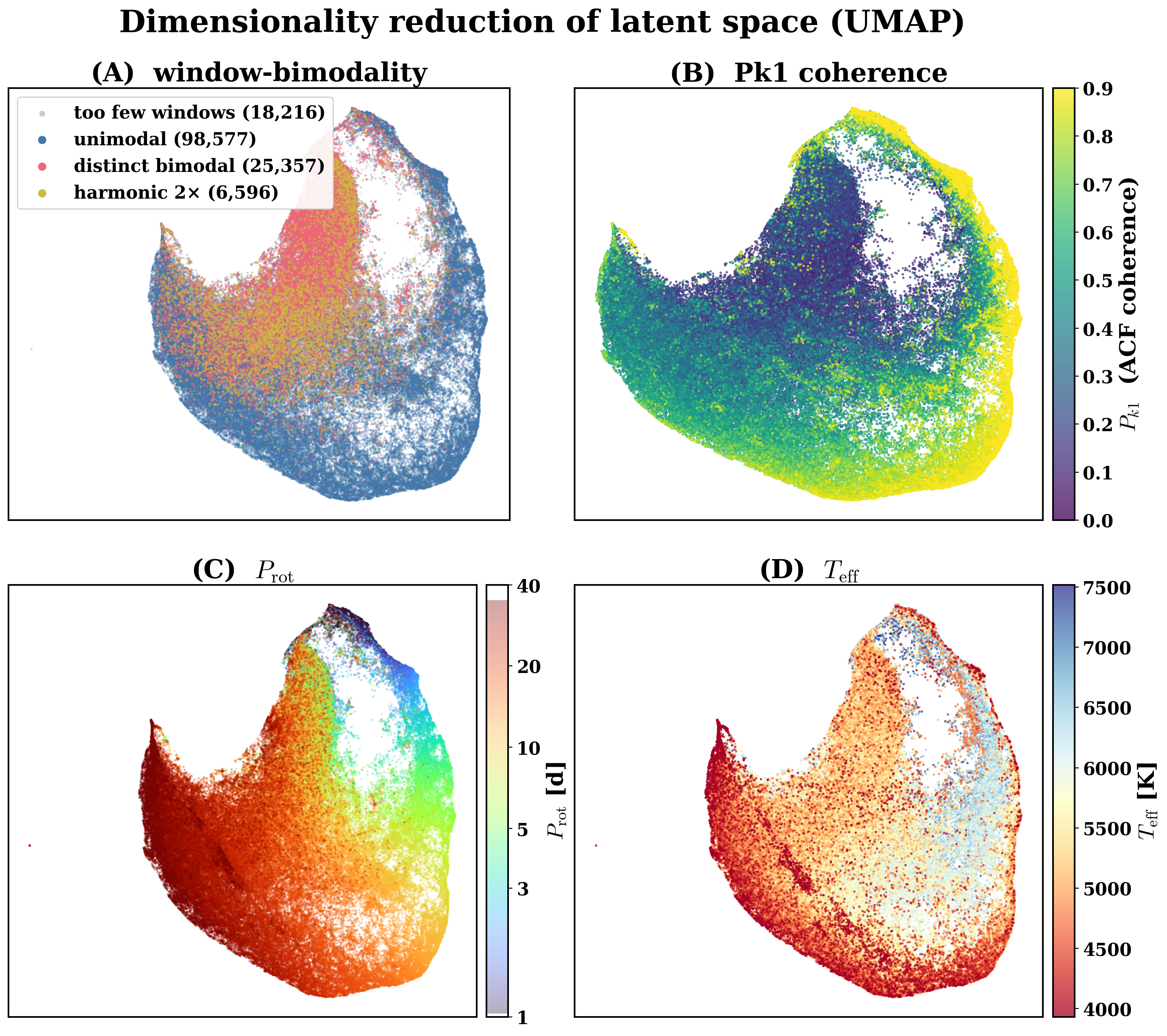}
    \caption{Dimensionality reduction of the $768$ latent space dimensions  of \emph{The Maunder} using UMAP. Each panel colors the two UMAP dimensions with values of a different property.}\label{fig:umap_2d}
\end{figure*}

\begin{figure*}
    \centering
    \includegraphics[width=0.7\linewidth]{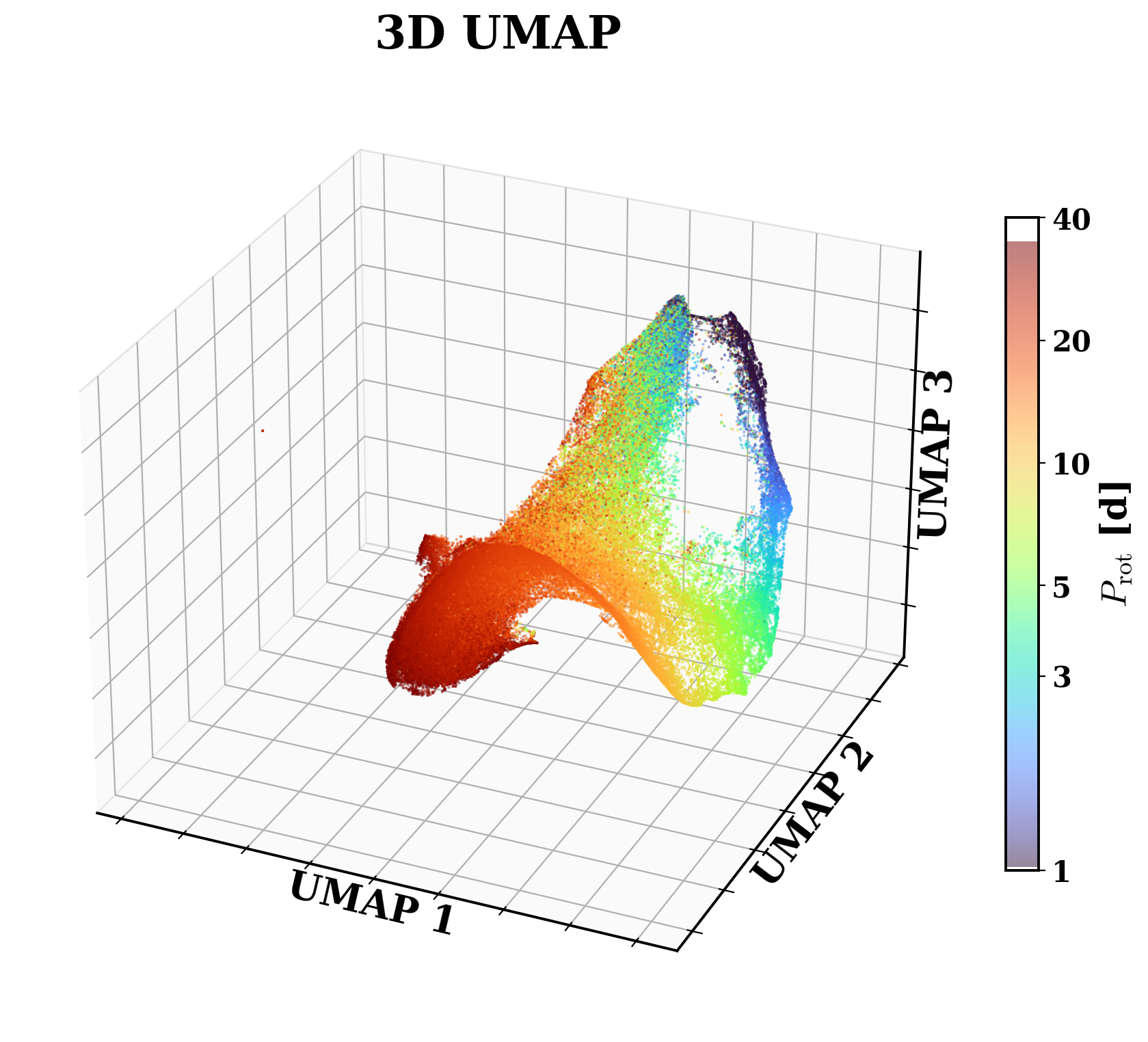}
    \caption{3D UMAP colored by $P_{\rm{rot}}$.}\label{fig:umap_3d}
\end{figure*}

\begin{figure*}
    \centering
    \includegraphics[width=0.7\linewidth]{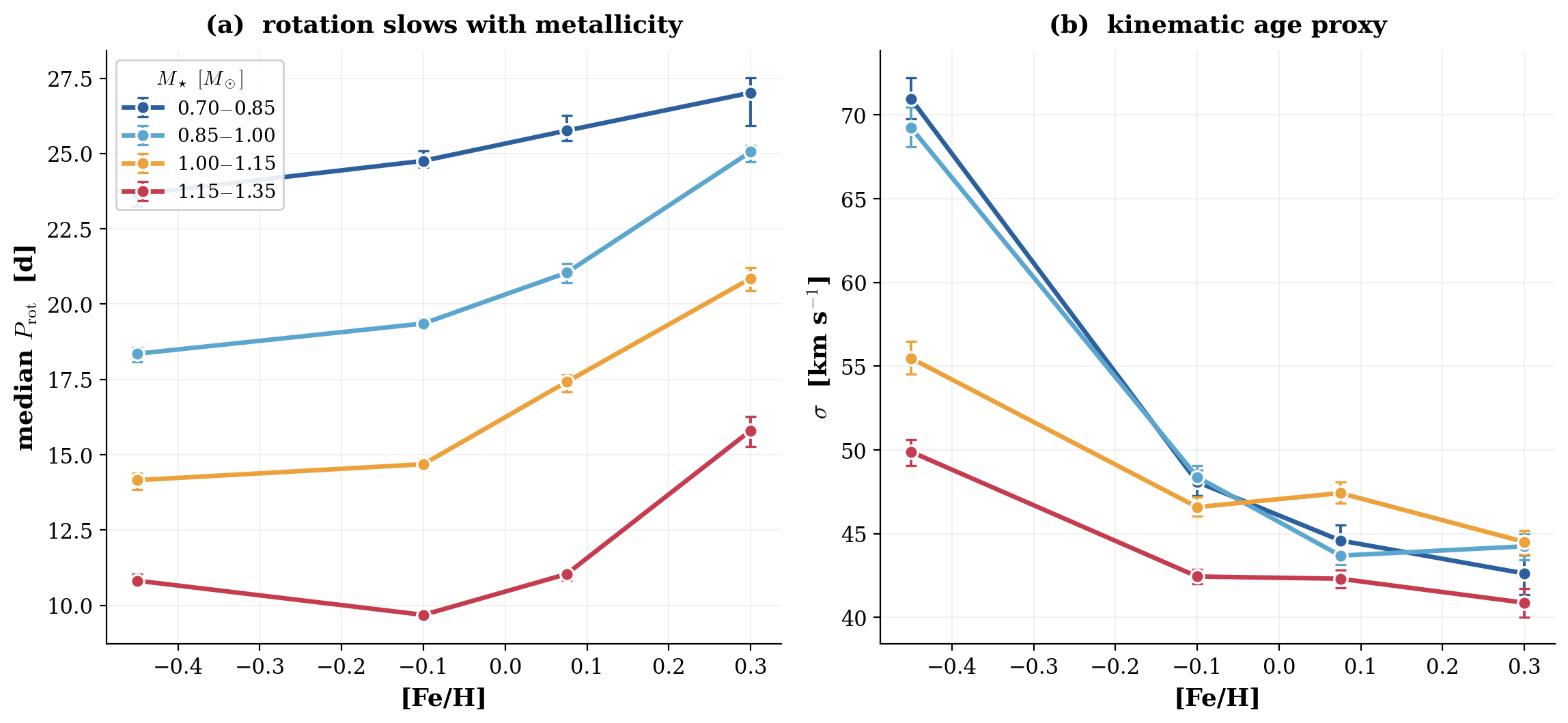}
    \caption{Difference between metallicity-period correlation (left panel) and metallicity-age correlation (right panel). Age is approximated through velocity dispersion - $\sigma$ }\label{fig:feh_prot_vdisp}
\end{figure*}

